\documentclass[preprint,subfigure,nofootinbib,aps,superscriptaddress,eqsecnum]{revtex4-1} %\documentclass[11pt, a4paper]{article}
\pdfoutput=1
\usepackage{extarrows}
\usepackage{graphicx}
\usepackage{subcaption}
\usepackage{amsmath}
\usepackage{amsfonts}
\usepackage{amssymb}
\usepackage{hyperref}
\usepackage{caption}
\usepackage{subcaption}
\usepackage{color}
\allowdisplaybreaks
\def\bea{\begin{eqnarray}}
\def\eea{\end{eqnarray}}
\def\be{\begin{equation}}
\def\ee{\end{equation}}

\begin{document}

	\title{Scalar dark matter and leptogenesis in the minimal scotogenic model}
	
	\author{Lavina Sarma}
	\email[Email Address: ]{lavina@tezu.ernet.in,sarmalavina@gmail.com}
	\affiliation{Department of Physics, Tezpur University, Assam-784028, India}
	\author{Pritam Das}
	\email[Email Address: ]{pritam@tezu.ernet.in,prtmdas9@gmail.com}
	\affiliation{Department of Physics, Tezpur University, Assam-784028, India}  
	\author{Mrinal Kumar Das}
	\email[Email Address: ]{mkdas@tezu.ernet.in,mkd.dlj@gmail.com}
	\affiliation{Department of Physics, Tezpur University, Assam-784028, India}

\begin{abstract}
	We study the minimal scotogenic model constituting an additional inert Higgs doublet and three sets of right-handed neutrinos. The scotogenic model connects dark matter, baryon asymmetry of the Universe and neutrino oscillation data. In our work, we obtain baryogenesis by the decay of TeV scale heavy neutral singlet fermion ($N_{2}$). We primarily focus on the intermediate-mass region of dark matter within $M_W<M_{DM}\le550$ GeV, where observed relic density is suppressed due to co-annihilation processes. We consider thermal as well as the non-thermal approach of dark matter production and explore the possibility of the lightest stable candidate being a dark matter candidate. Within the inert Higgs doublet (IHD) desert, we explore a new allowed region of dark matter masses for the non-thermal generation of dark matter with a mass splitting of 10 GeV among the inert scalars. We also see the variation of relic abundance for unequal mass splitting among the scalars. The KamLand-Zen bound on the effective mass of the active neutrinos is also verified in this study.% keeping intact constraints from Planck limit as well as direct detection experiment XENON1T. }
\end{abstract}


%%%%%%%%%%%%%%%%%%%%%%%%%%%%%%%%%

	\maketitle
	\flushbottom 
\section{Introduction}
The Standard Model (SM) of particle physics is an affluent and self-consistent one in the current scenario. However, it is not accountable for explaining various problems persisting in the Universe. Among all the anomalies, baryon asymmetry of the Universe \cite{leptogenesis,Hugle:2018qbw}, absolute neutrino mass \cite{deSalas:2017kay},  dark matter \cite{bertone2005particle,Moore:1999nt} are the ones drawing much of the attention in the ongoing studies nowadays.

Successively, there has been significant growth in the past few years in providing pieces of evidence to these mysterious and yet interesting form of non-baryonic matter, commonly termed as dark matter (DM) in the present Universe. 
The significant lines of evidence of DM include observations in galaxy cluster by Fritz Zwicky \cite{Zwicky:1933gu} in 1933, gravitational lensing (which could allow galaxy cluster to act as gravitational lenses as postulated by Zwicky in 1937) \cite{Treu:2012sn}, galaxy rotation curves in 1970 \cite{Rubin:1970zza}, cosmic microwave background \cite{cosmicmicrowave} and the most recent cosmology data given by Planck satellite \cite{Ade:2015fva} are some of the most remarkable ones. From the recent Planck satellite data, it is certain that approximately $27\%$ of the present Universe is comprised of DM, which is about five times more than the baryonic matter. A brief discussion about the conditions required to be fulfilled by particle candidates for DM is found in this paper \cite{Taoso:2007qk}, from which it is confirmed that the possibility of SM particle to be a DM candidate is ruled out. This has resulted in the extension of the SM, of which the weakly interacting massive particle (WIMP) paradigm is the most discussed framework. 

A notable co-occurrence frequently termed as the WIMP miracle \cite{Kolb:1990vq} is feasible in the WIMP paradigm, where a dark matter candidate typically with an electroweak scale mass and electroweak alike interactions can produce correct dark matter relic abundance. WIMPs can be thermally produced in the early Universe as the interactions governing them are of electroweak scale. Thus, relic abundance of a thermal DM candidate can be generated while the interactions freeze out, ensuing the expansion as well as the cooling of the Universe. Also, the WIMP paradigm foretells the observable DM nucleon scattering cross-section through the same interactions that were operational at the time of freeze-out. However, many dark matter direct detection experiments like LUX \cite{Akerib:2016vxi}, PandaX-II \cite{Tan:2016zwf}, and XENON1T \cite{Aprile:2017iyp} have reported their null results. Therefore, the exclusion curve in the mass-cross section plane is lowered. Similar null results have been obtained from the Large Hadron Collider(LHC), which further gives an upper bound on the DM interaction with the SM particles. A strict constraint on the WIMP parameter space can be summarized from the different null results.

Besides DM, the baryon asymmetry of the Universe is another puzzle, which is the observed imbalance in the baryonic matter and anti-baryonic matter in the observable Universe. 
%There are a set of conditions that were inspired by the recent discoveries of the cosmic microwave background \cite{cosmicmicrowave} and CP violation \cite{CPviolation} in the neutral kaon system which are necessary for a baryon generating interaction to produce matter and antimatter at different rates.
A particle to create baryon asymmetry, it must satisfy the Sakharov conditions \cite{Sakharov:1967dj}, which demands baryon number (B) violation, C and CP violation, and departure from thermal equilibrium. As these conditions cannot be fulfilled within the SM in an adequate amount, we need formalism beyond the SM. Of these criteria, the out-of-equilibrium decay of a heavy particle leading to the generation of the baryon asymmetry of the Universe (BAU) has so far been a widely known mechanism for baryogenesis \cite{Weinberg:1979bt}. We can incorporate such a mechanism via leptogenesis \cite{leptogenesis}, where a net leptonic asymmetry is generated first, which further gets converted into baryogenesis through $(B+L)$ violating electroweak sphaleron phase transitions \cite{EWSphaleron}.
% In this kind of thermal leptogenesis, the RHNs are produced in the thermal bath through scattering, much before their CP-violating out-of-equilibrium decays give rise to ancient lepton asymmetry. Also, an advantage of this outline is that the CP-violating out-of-equilibrium decays of the same heavy fields that generate the lepton asymmetry is responsible for explaining the origin of the light neutrino masses \cite{neutrinomasspdg} in the seesaw mechanism\  \cite{MINKOWSKI1977421,seesawRN,seesaw11,seesaw3,seesaw4}. This standard formulation can be closely related to type-I seesaw mechanism \cite{MINKOWSKI1977421}. However, there is a fundamental limitation of the thermal leptogenesis as it requires very heavy RHN mass.
A rich literature is available for various leptogenesis processes \cite{neutrinomasspdg,Minkowski,Mohapatra,Yanagida:1979as,Schechter:1980gr,Glashow:1979nm}.  In the case of an elementary scenario, mostly referred to as vanilla leptogenesis, where the lower mass bound, by the allowance of flavor effect, comes down to be about $M_{1}^{min}= 10^{8}$ GeV \cite{Hugle:2018qbw,Blanchet:2008pw}.
Owing to the fact that the CP asymmetry in RHN decays is a consequence of the active and sterile neutrino masses along with the necessity of tiny SM neutrino masses, the high mass scale of RHN is needed \cite{Davidson:2002qv,Buchmuller:2004nz}. %Nevertheless, such a high mass scale of RHN is disagreeable for several reasons, of which, the mere possibility of detecting the dynamics of leptogenesis in future collider experiments \cite{ejchun} is of much significance. Another reason is that the high-scale leptogenesis may be precluded due to the future detection of lepton number violation at low energies \cite{hugle4}. Thus, these observations act as a catalyst to opt for other alternatives to the archetype of standard thermal leptogenesis in the type-I seesaw mechanism that copes up to produce the BAU at much lower RHN mass order.
Our work is carried out depending on the idea of low mass RHN as mentioned in \cite{Hugle:2018qbw}, the study of thermal leptogenesis in Ernest Ma's scotogenic model \cite{Ma:2006km,Ma:2017kgb}, which is considered to be the simplest model of radiative neutrino masses. Hence, we choose the scale of the RHN in such a fashion that it can satisfy the observed value of BAU and does not bother the dark matter phenomenology. % The scotogenic model is of much significance as we can relate the light SM neutrino mass with the physics of dark matter \cite{Ma:2006km}. %However, we have implemented the scotogenic model and not the pure IHDM, and because of the choice of RHN mass viz. quite heavier than the DM mass, it implies no effect on the relic generated for the DM candidate. 

This work primarily focuses on the IHDM desert, i.e., $M_{W} < M_{DM} \le 550$ GeV, wherein the generation of the relic abundance is prohibited as mentioned in various literatures\cite{Ma:2006km,LopezHonorez:2006gr,Ahriche:2017iar}. The core reason behind this discrepancy is that in the IHDM desert, the annihilation cross-section of the dark matter is large compared to the amount necessary to produce the correct relic abundance via the freeze-out mechanism. Thus, we get an underabundant DM in this regime due to the large annihilation rates. Though the lower bound of the IHDM desert is rigid, the upper bound can be a little flexible depending on the choice of parameters such as the DM-Higgs coupling and the mass splitting between the inert scalars. Thus, we try to see the viability of IHDM desert, concentrating on the upper bound satisfying the relic abundance value with latest restrictions from direct detection experiment XENON1T \cite{Aprile:2017iyp}. The production of a correct relic in this regime can be possible by fine-tuning of the DM-Higgs coupling and suitable mass splitting of the other inert scalars. 

Motivated by these factors, in this model, the SM is extended by a Higgs doublet field ($\eta$) and three singlet neutral fermions ($N_{k}$), which are odd under $Z_{2}$ symmetry, in contradiction to the SM particles which are $Z_{2}$ even. The possibility of a DM candidate comes from the $Z_{2}$ odd lightest particle. Whereas, leptogenesis is a result of the $Z_{2}$ odd fermions, i.e., the heavy RHN, which occurs via the out-of-equilibrium decay into the SM leptons and the inert Higgs doublet \cite{LopezHonorez:2006gr}. The entire work is carried out keeping the dark matter mass in the intermediate dark matter mass range, also known as IHDM desert, which lies between $M_{W} < M_{DM} \le 550$ GeV. Leptogenesis is obtained for this very range of dark matter mass with the decay of $N_{2}$ which is the next to lightest RHN. Also, an important criterion that is kept intact is the sum of neutrino masses and its effective mass being consistent with the constraints from Planck data and neutrinoless double beta decay experiment, KamLAND-Zen. We also check the relic abundance of the dark matter candidate (lightest of $\eta$) for different choices of mass splitting between the scalars of the inert scalar doublet. We further investigate the parameter space, i.e. the values of DM-Higgs coupling and dark matter mass for which it satisfies the bounds from relic abundance and direct detection experiment. Furthermore, we also study the mixture of thermal and non-thermal production of DM abundance for various masses within the IHDM desert. In one of the cases, we have considered mass splitting of the scalars in the inert doublet to be 10 GeV and studied the criteria that satisfy the observed relic for higher DM masses within the IHDM desert via purely thermal production as well as non-thermal production. The non-thermal production process is solely to enhance the relic of DM which is under-abundant in the IHDM desert, produced via thermal mechanism. This can be made possible by the late decay of the RHN, $N_{1}$ into DM and SM leptons for very small decay width of $N_{1}$, which further makes it incompetent to produce the BAU.
% It can be noted that in non-thermal production, relic can be obtained for DM masses within the IHDM desert by appropriate choice of decay width and initial abundance of DM. We, therefore, can bring down a preferred range of the above-mentioned parameters to generate the correct relic abundance in the non-thermal production of DM.  \\

The rest of the paper is divided into six sections, where section\eqref{eq:6} includes a brief introduction of the scotogenic model involving the generation of neutrino mass. Section\eqref{eq:8} and section\eqref{eq:NDBD} constitutes discussions on baryogenesis in scotogenic model and neutrinoless double beta decay, respectively. Thermal and non-thermal production of dark matter is discussed in section\eqref{eq:7}. A detailed numerical analysis, along with results, are shown in section\eqref{eq:9} followed by the conclusion given in section\eqref{eq:10}.
	
\section{Scotogenic model}\label{eq:6}

 Scotogenic model is an extension of the IHDM \cite{LopezHonorez:2006gr} and the IHDM is nothing but a minimal extension of the SM by a Higgs field which is a doublet under $SU(2)_{L}$ gauge symmetry with hypercharge $Y=1$ and a built-in discrete $Z_{2}$ symmetry \cite{LopezHonorez:2006gr,Ahriche:2017iar,Deshpande:1977rw,Cirelli:2005uq,Barbieri:2006dq,Ma:2006wm,Hambye:2009pw,Dolle:2009fn,Honorez:2010re,Gustafsson:2012aj,Borah:2017dfn,Goudelis:2013uca,Arhrib:2013ela,Bhattacharya:2019fgs,Borah:2019aeq}. The necessity of this modification took place as the IHDM could only accommodate dark matter, whereas it failed in explaining the origin of neutrino masses at a renormalizable level \cite{Borah:2017dfn}. In this model, three neutral singlet fermions $N_{i}$ with $i=1,2,3$ are added in order to generate neutrino masses and assign them with a discrete $Z_{2}$ symmetry. In view of $N_i$, the neutrinos can get masses in two ways. One of the ways is similar to the type-I seesaw mechanism \cite{Minkowski,Mohapatra,Schechter:1980gr,Glashow:1979nm}, where the neutrino masses arise as a result of $N_i$ being $Z_{2}$ even. Also, it is limited to show no dark matter phenomenology of the IHDM and keeps the neutrino masses decoupled from the DM characteristics. Therefore, we opt for the other way in which $N_i$ is odd under $Z_{2}$ symmetry, whereas the SM fields remain $Z_{2}$ even. Symbolic transformation of the particles under $Z_{2}$ symmetry is given by,
	 
	 \begin{equation}
	 N_{i}\longrightarrow -N_{i},~ \eta\longrightarrow -\eta,~ \Phi\longrightarrow \Phi,~\Psi\longrightarrow \Psi,
	 \end{equation}
	 where $\eta$ is the inert Higgs doublet, $\Phi$ is the SM Higgs doublet and $\Psi$ denotes the SM fermions.
	 The new leptonic and scalar particle content can thereafter be represented as follows under the group of symmetries $SU(2) \times U(1)_{Y} \times Z_{2}$:
	  \begin{equation*}
	 \begin{pmatrix}
	 \nu_{\alpha}\\
	 l_{\alpha}
	 \end{pmatrix}_{L} \sim (2,  -\dfrac{1}{2}, +) ,~ l^{c}_{\alpha} \sim (1,1,+), ~
	 \begin{pmatrix}
	 \Phi^{+} \\ 
	 \Phi^{0}\\ 
	 \end{pmatrix} \sim (2,\frac{1}{2},+), 
	 \end{equation*}
	 \begin{equation}
	 	 N_{i} \sim (1,1,-), ~
	 \begin{pmatrix}
	 \eta^{+}\\
	 \eta^{0}\\
	 \end{pmatrix} \sim (2,1/2, -).
	 \end{equation}
	 The scalar doublets are written as follows :
	 \begin{equation}
	 \eta=
	 \begin{pmatrix}
	 \eta^{\pm}\\
	 \frac{1}{\sqrt{2}}(\eta^{0}_{R} + i\eta^{0}_{I}) \\
	 \end{pmatrix}, \quad
	 \Phi=
	 \begin{pmatrix}
	 \Phi^{+}\\
	\frac{1}{\sqrt{2}} (h+i\xi)
	 \end{pmatrix}.
	 \end{equation} 
%	 \hspace{3cm}
	 \begin{figure}[h]
\centering
	 	\includegraphics[width=0.4\textwidth]{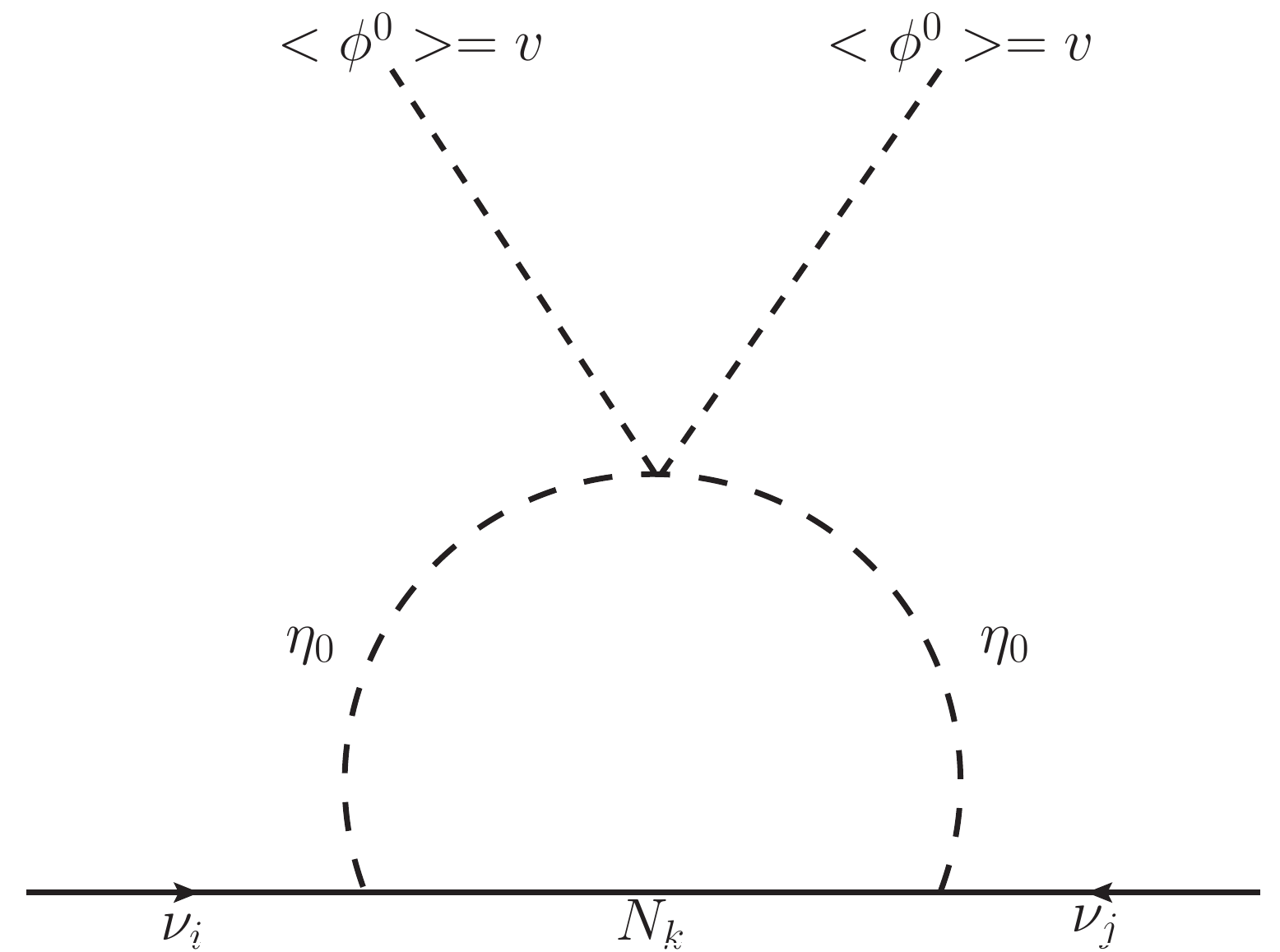}
	\caption{ One- loop contribution of neutrino mass generation with the exchange of right handed neutrino $N_k$ and the scalar $\eta_0$. } \label{fig1}
 \end{figure}
  We have no Dirac mass term with $\nu$ and $N$; however, the similar Yukawa-like coupling involving $\eta$ is allowed. Nevertheless, the scalar cannot get a VEV. The neutrino mass can be generated through a one-loop mechanism, which is based on the exchange of $\eta$ particle and a heavy neutrino. In fig \ref{fig1}, we see two Higgs fields $\phi^{0}$ are involved. They will not propagate but will acquire VEV after the EWSB.
  
The lagrangian involving the newly added field is :
  	\begin{equation}
  	\mathcal{L} \supset \frac{1}{2}(M_{N})_{ij}N_{i}N_{j} + Y_{ij}\bar{L_{i}}
  	\tilde{\eta}N_{j} + h.c \label{lag1}
  	\end{equation}
  where, the $1^{st}$ term is the Majorana mass term for the neutrino singlet and the $2^{nd}$ term is the Yukawa interactions of the lepton.
  The new potential on addition of the new inert scalar doublet is:
  	\begin{equation}
 	\begin{split}
	V_{Scalar} = &m_{1}^{2}\Phi^{+}\Phi + m_{2}^{2}\eta^{+}\eta + \frac{1}{2}\lambda_{1}(\Phi^{+}\Phi)^{2} + \frac{1}{2}\lambda_{2}(\eta^{+}\eta)^{2}+ \lambda_{3}(\Phi^{+}\Phi)(\eta^{+}\eta) \\&+
  \lambda_{4}(\Phi^{+}\eta)(\eta^{+}\Phi) 
  +\big[\frac{\lambda_{5}}{2}(\Phi^{+}\eta)^{2} + h.c.\big]\label{eqp1}
    	\end{split}
  	\end{equation}
  All the parameters in Eq. \eqref{eqp1} are real by hermicity of the Lagrangian, except for $\lambda_{5}$. Since, the bilinear term $(\Phi^{+}\eta)$ is forbidden by the exact $Z_{2}$ symmetry, therefore one can always choose $\lambda_{5}$ real by rotating the relative phase between $\Phi$ and $\eta$. Furthermore, after the spontaneous symmetry breaking like in the SM, we are left with one physical Higgs boson \textit{h} which resembles the SM Higgs boson, as well as four dark scalars: one CP even($\eta^{0}_{R}$), one CP odd($\eta_{I}^{0}$) and a pair of charged ones ($\eta^{\pm}$). The masses of these physical scalars are:
  	\begin{equation}
  	\begin{split}
    m^{2}_h =& -m^{2}_{1} = 2\lambda_{1}\textit{v}^{2},\\
  m^{2}_{\eta^{\pm}} =& m^{2}_{2}+\frac{1}{2}\lambda_{3}\textit{v}^{2},\\
  m^{2}_{\eta_{R}^{0}} =& m^{2}_{2} + \frac{1}{2}(\lambda_{3}+\lambda_{4}+\lambda_{5})\textit{v}^{2},\\
  m^{2}_{\eta_{I}^{0}} = &m^{2}_{2} + \frac{1}{2}(\lambda_{3}+\lambda_{4}-\lambda_{5})\textit{v}^{2}.
    	\end{split}\label{mas1}
    	\end{equation}
  It is clear from the above equations that all the scalar couplings are written in terms of physical scalar masses and $m_{2}$ , thereby providing six independent parameters of the model to be :
  	$ \{m_{2},m_{\textit{h}},m_{\eta_{R}^{0}},m_{\eta_{I}^{0}},m_{\eta^{\pm}},\lambda_{2} \}{\large {\tiny }} $.
Here, $m_{\textit{h}}$ is the mass of SM-Higgs, $m_{\eta_{R}^{0}}$, $m_{\eta_{I}^{0}}$ and $m_{\eta^{\pm}}$ are the masses of CP-even, CP-odd and charged scalars of the inert doublet respectively. In this work, as we have considered the CP-even scalar to be the lightest particle and a probable DM candidate, so we consider $\lambda_{5} < 0$ without any loss of generality. Moreover, the mass difference between the real and imaginary component of the inert doublet $\eta$ from \eqref{mas1} can be written as , $m_{\eta_{R}^0}^2-m^2_{\eta_{I}^0}=\lambda_{5}v^2$. Therefore, in the limit $\lambda_{5}\rightarrow 0$ leads to the mass degeneracy of the neutral components of the inert doublet. Again, the case of vanishing $\lambda_5$ would lead to vanishing neutrino mass, as the $\lambda_5$ in \eqref{eqp1} associate with the lepton number violation term in \eqref{lag1}. Therefore, considering $\lambda_5\rightarrow0$ allows us to recover the lepton number global symmetry, and following the 't Hooft scenario \cite{tHooft:1980xss}, the smallness of $\lambda_{5}$ is essential to obtain the lepton asymmetry, which would have been lost if considered to be zero, is acceptably natural. Throughout our analysis, for the coupling constants we follow the inequality relation $\lambda_4>>\lambda_5$.
We have a simplified diagram that can be split further into two diagrams and from which the mass can be easily calculated by considering mechanism after EWSB. 
   \begin{figure}[h]
  %\begin{center}
  \centering
  	\includegraphics[width=0.5\textwidth]{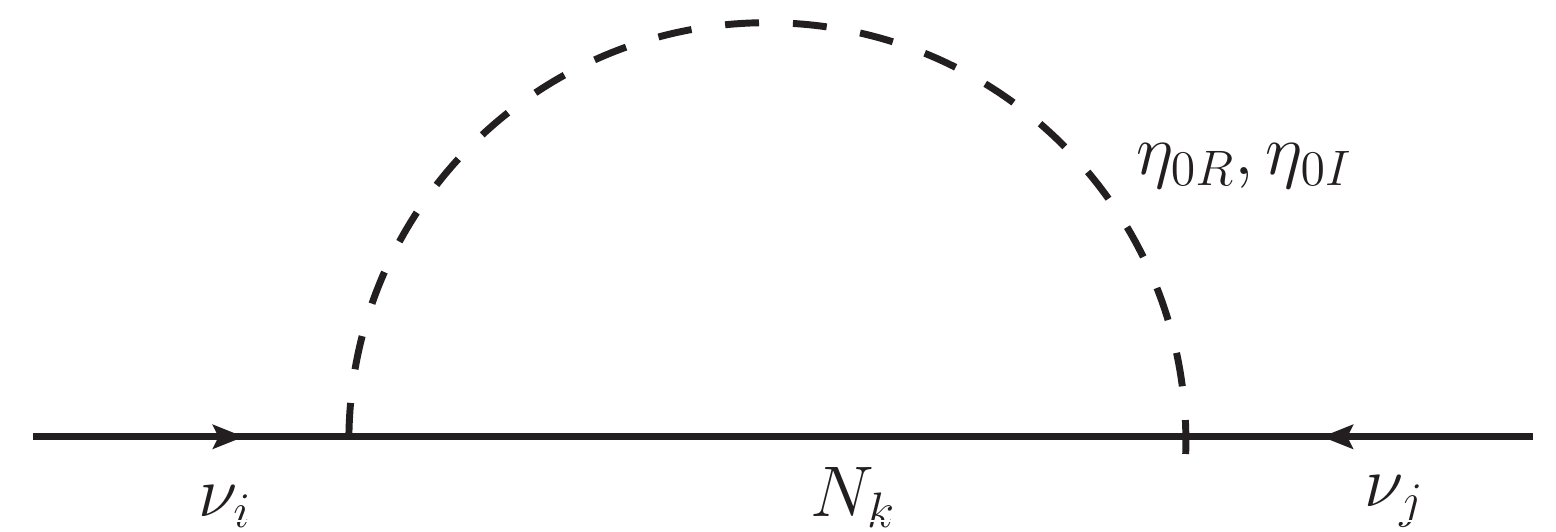}
  	\caption{ One-loop diagram with exchange of $\eta_{R}^{0}$ and $\eta_{I}^{0}$. $\nu_i$ and $\nu_j$ representing two different generations of active neutrinos. $N_k$ is the right handed neutrino.}\label{nfig2}
%  \end{center}
 \end{figure} 
 Calculation on the basis of one diagram is sufficient and considered as other would be same except for $\eta_{R}^{0}$ replaced by $\eta_{I}^{0}$. The neutrino mass matrix arising from the radiative mass model is given by \cite{Ma:2006wm,Merle:2015ica}:
 \begin{equation}
 \begin{split}
 \textit{M}_{ij}^{\nu}=&\sum_{k} \frac{h_{ik}h_{jk}}{16\pi^{2}}M_{\textit{k}}\left[\frac{m_{\eta_{R}^{0}}^{2}}{m_{\eta_{R}^{0}}^{2}-M_{\textit{k}}^{2}}\:ln\frac{m_{\eta_{R}^{0}}^{2}}{M_{\textit{k}}^{2}}-\frac{m_{\eta_{I}^{0}}^{2}}{m_{\eta_{I}^{0}}^{2}-M_{\textit{k}}^{2}}\:ln\frac{m_{\eta_{I}^{0}}^{2}}{M_{\textit{k}}^{2}}\right]\\
 \equiv& \sum_{k} \frac{h_{ik}h_{jk}}{16\pi^{2}}M_{\textit{k}}[L_{k}(m^{2}_{\eta_{R}^{0}}) - L_{k}(m^{2}_{\eta_{I}^{0}})],\label{eq7}
 \end{split}
 \end{equation}
 where $M_{k}$ represents the mass eigenvalue of the mass eigenstate $N_{k}$ of the neutral singlet fermion $N_{k}$ in the internal line with indices j=1,2,3 running over the three neutrino generation with three copies of $N_k$. The function $L_{k}(m^{2})$ used in Eq. \eqref{eq7} is given by:
 \begin{equation}
 	L_{k}(m^{2})= \frac{m^{2}}{m^{2}-M^{2}_{k}}\ln\frac{m^{2}}{M^{2}_{k}}
 \end{equation}
 In our study, we calculate the Yukawa couplings by the incorporation of the constraints on the sum of neutrino masses \cite{Aghanim:2018eyx} and the neutrino oscillation data \cite{deSalas:2017kay}. For simplicity of the Yukawa coupling calculation, we write the mass formula given by Eq. \eqref{eq7}, in the form similar to type-I seesaw formula\cite{Mahanta:2019gfe}:
 \begin{equation}
 	M_{\nu}= Y\Lambda^{-1}Y^{T},
 \end{equation}
 where $\Lambda$ is a diagonal matrix represented by\cite{Takashi}:
 \begin{equation}
 	\Lambda_{k}= \frac{M_{k}}{16\pi^{2}}\left[\frac{m_{\eta_{R}^{0}}^{2}}{m_{\eta_{R}^{0}}^{2}-M_{\textit{k}}^{2}}\:ln\frac{m_{\eta_{R}^{0}}^{2}}{M_{\textit{k}}^{2}}-\frac{m_{\eta_{I}^{0}}^{2}}{m_{\eta_{I}^{0}}^{2}-M_{\textit{k}}^{2}}\:ln\frac{m_{\eta_{I}^{0}}^{2}}{M_{\textit{k}}^{2}}\right].
 \end{equation}
 
The light neutrino mass matrix \eqref{eq7} can be diagonalised by an unitary matrix known as the {\it  Pontecorvo-Maki-Nakagawa-Sakata(PMNS)} matrix.

The diagonal light neutrino mass matrix can be written as:
\begin{equation}
	M^{diag}_{\nu}= U^{\dagger}M_{\nu}U^{*}
\end{equation} 
Also, we use a special yet one of the most popular types of parametrization known as the {\it Casas-Ibarra parametrization }\cite{Casas:2001sr} in order to link the Yukawa coupling with the light neutrino parameters.
\begin{equation}
	Y= U\sqrt{M^{diag}_{\nu}}R^{\dagger}\sqrt{\Lambda},\label{eq:CI}
\end{equation}
where $R$ is a complex orthogonal matrix satisfying the condition $R^{T}R=1$. We also parameterized the $R$ matrix as per our convenience and the orthogonal complex matrix $R$ takes the form,
\begin{equation}\label{eq:11}
R=
\begin{pmatrix}
0 & \cos Z & \sin Z\\
0 & -\sin Z & \cos Z \\
1 & 0 & 0\\
\end{pmatrix},
\end{equation}
where, $Z= (z_{R}+ iz_{I})$ with $z_{R},z_{I}\in [0,2\pi]$ \cite{Ibarra:2003up}. In our case, we consider the values 1.42 and 1.6232 respectively for normal hierarchy(NH). In the case of inverted hierarchy, we arbitrarily choose lower values of $z_{R}$= 0.22 and $z_{I}$= 0.58, which contributes to a slight difference in the baryogenesis plot as a function of RHN $N_{2}$.  This choice of the orthogonal matrix $R$ is made to calculate the Yukawa couplings related by the {\it Casas- Ibarra parametrization} given in Eq.(\ref{eq:CI}), in order to obtain a non-zero complex term for $(Y^{\dagger}Y)_{22}$ which is inversely proportional to the CP asymmetry $\epsilon_{2}$. Since $\epsilon_{2}$ is directly dependent on $(Y^{\dagger}Y)_{21}$ and $(Y^{\dagger}Y)_{23}$ as well, the requirement of these quantities to be non-zero is a must. Therefore, such a choice of $R$ as in Eq.\eqref{eq:11} is adequate in fulfilling the foresaid criteria.  The evaluated Yukawa matrix in NH mode from \eqref{eq:CI} is given by,
\begin{equation}\label{eq:YU}
	Y_{NH}=\left(\begin{array}{ccc}
		-9.27224\times 10^{-6}+0.0000412i & -0.0013963-0.00034838i & -0.0443544\\
		-0.00001429+0.00004271i & -0.0014436-0.0005405i & -0.022009+0.004795i\\
		5.60918\times 10^{-6}-0.00004588i & 0.001558+0.000227i & 0.020248+0.00483i
	\end{array}\right)
\end{equation}
and that for IH mode is given by:
\begin{equation}\label{eq:YUK}
Y_{IH}=\left(\begin{array}{ccc}
-0.00002917+3.70529\times 10^{-6}i & 0.00010793+0.00024547i & 0.0743562\\
-0.000025069+6.74852\times 10^{-6}i & 0.000070818+0.00021745i & -0.0409688+0.0062949i\\
0.000031937-6.25485\times 10^{-6}i & -0.00013739-0.0002630i & 0.0367029+0.0055138i
\end{array}\right)
\end{equation}
Also, for the Yukawa coupling values obtained in this work, the lepton flavor violating process  $l_{\alpha}\rightarrow l_{\beta}\gamma$ is possible. Bounds from various LFV processes in this model are discussed in the following subsection. 
\subsection{Bounds on this model}

	\subsubsection{Lepton flavor violating processes}
It is well known that lepton flavor violating processes put significant bound on the model parameter space. The size of the LFV is controlled by the
lepton number violating Yukawa couplings $Y_{ij}$.
The LFV processes such as  $l_{\alpha}\rightarrow l_{\beta}\gamma,~l_{\alpha}\rightarrow 3 l_{\beta}$ and $\mu-e$ conversion in nuclei within the framework of scotogenic model put significant bounds \cite{Takashi}.\\
In case of radiative lepton decay, the branching ratio of $l_{\alpha}\rightarrow l_{\beta}\gamma$ is given by-
\begin{equation}
\text{Br}(l_{\alpha}\rightarrow l_{\beta}\gamma)=\frac{3(4\pi^{3})\alpha_{em}}{4G_{F}^{2}}|A_{D}|^{2} \text{Br}(l_{\alpha} \rightarrow l_{\beta}\nu_{\alpha}\bar{\nu{\beta}})
\end{equation}
For three body decay process like  $l_{\alpha}\rightarrow 3 l_{\beta}$, the branching ratio is given by-
\begin{equation}
\begin{aligned}
\text{Br}(l_{\alpha}\rightarrow 3 l_{\beta})=\frac{3(4\pi^{2})\alpha_{em}^{2}}{8G_{F}^{2}} \bigg[|A_{ND}|^{2}+|A_{D}|^{2}\bigg(\frac{16}{3}log\bigg(\frac{m_{\alpha}}{m_{\beta}}\bigg)-\frac{22}{3}\bigg)\\  +\frac{1}{6}|B|^{2}+\bigg(-2A_{ND} A_{D}^{*}+\frac{1}{3}A_{ND} B^{*}-\frac{2}{3}A_{D}B^{*}+h.c\bigg)\bigg] \\ 
\times \text{Br}(l_{\alpha} \rightarrow l_{\beta}\nu_{\alpha}\bar{\nu{\beta}})
\end{aligned}
\end{equation}
The conversion rate, normalized
to the the muon capture rate, can be expressed as -
\begin{equation}
\begin{aligned}
\text{CR}(\mu-e,Nucleus)=\frac{p_{e}E_{e}m_{\mu}^{3}G_{F}^{2}\alpha_{em}^{3}Z_{eff}^{4}F_{p}^{2}}{8\pi^{2}Z \Gamma_{capt}}\times \bigg[|(Z+N)(g_{LV}^{(0)}+g_{LS}^{(0)})+(Z-N)(g_{LV}^{(1)}+g_{LS}^{(1)})|^{2}\\  +|(Z+N)(g_{RV}^{(0)}+g_{RS}^{(0)})+(Z-N)(g_{RV}^{(1)}+g_{RS}^{(1)})|^{2}\bigg] 
\end{aligned}
\end{equation}
The notations we have used in the above mentioned relations are explicitly taken from \cite{Takashi}. The MEG collaboration has been able to set the impressive bound on muon decay  Br$(l_{\alpha} \rightarrow l_{\beta}\gamma)<4.2 \times 10^{-13}$\cite{TheMEG}. In case of $l_{\alpha}\rightarrow 3 l_{\beta}$ decay contraints comes from SINDRUM experiment to be $\text{Br}(l_{\alpha} \rightarrow 3l_{\beta})<10^{-12}$ which has been set long ago.
In our analysis, for NH, we obtain:
Br$(\mu \rightarrow e\gamma)= 6.22\times 10^{-18}$, Br($\mu \rightarrow 3e$)= $7.31839\times 10^{-34}$, CR($\mu\rightarrow e$)= $2.64278\times 10^{-36}$. Similarly for IH, we obtain: Br($\mu \rightarrow e\gamma$)= $8.36708\times 10^{-19}$, Br($\mu \rightarrow 3e$)= $3.60477\times 10^{-34}$, CR($\mu\rightarrow e$)= $9.59989\times 10^{-37}$. For Yukawa coupling values less than $10^{-4}$, as required by neutrino mass constraints, one can get branching ratio value for the process $l_{\alpha}\rightarrow l_{\beta}\gamma$ below the experimental bound given by MEG collaboration\cite{BorahBR}.  However, we have not considered LFV processes related to $\tau$ lepton (such as $\tau\rightarrow e\gamma,~\tau\rightarrow \mu\gamma,~\tau\rightarrow3e,~\tau\rightarrow3\mu$) in our study, as they are less sensitive to experiments (exceptions in case of high-luminosity electron-positron collider experiments like SuperB, Belle II may improve the bound, which we have not considered in our study). Hence, they do not have robust bounds on our model parameter spaces. 
\subsubsection{Stability constraints}
The stability of the scalar potential demands that the potential should be bounded from below, {\it i.e.}, it should not approach negative infinity along any direction of the field space at large field values. With large field the quadratic terms of
the scalar potential in eqn. \eqref{eqp1} are smaller compared to the quartic terms.
This scalar potential will be bounded from below if the following conditions are satisfied \cite{Deshpande:1977rw},
\begin{equation*}
	\lambda_1(\Lambda)\ge0;~~~\lambda_2(\Lambda)\ge0;~~~\lambda_{3}(\Lambda)\ge-2\sqrt{\lambda_{1}(\Lambda)\lambda_{2}(\Lambda)}~~\text{and}~~\lambda_{L,S}(\Lambda)\ge-\sqrt{\lambda_1(\Lambda)\lambda_2(\Lambda)}.
\end{equation*}
Here, $\lambda_{L,S}=\frac{1}{2}(\lambda_3+\lambda_4\pm\lambda_{5})$. The coupling constants are evaluated at a scale $\Lambda$ using RG equations. 
\subsubsection{Perturbativity constraints}
For IDM to behave as a perturbative quantum field theory at a given scale $\Lambda$, one must impose the condition on the couplings of the potential \ref{eqp1}, and they are as follows \cite{Lee:1977eg},
\begin{equation}
	|\lambda_1(\Lambda),~\lambda_2(\Lambda),~\lambda_3(\Lambda),~\lambda_4(\Lambda),~\lambda_5(\Lambda)|\le4\pi.
\end{equation}
\subsubsection{Unitarity bounds}
Unitarity bounds on the couplings are evaluated by considering scalar-scalar, gauge boson-gauge boson, and scalar-gauge boson scatterings \cite{Lee:1977eg}. In general, unitarity bounds are the couplings of the physical bases of the scalar potential. Nevertheless, the couplings for the scalars are quite complicated, therefore we consider the couplings of the non-physical bases before EWSB. Then the S-matrix, which is expressed in terms of the non-physical fields, is transformed into an S-matrix for the physical
fields by making a unitary transformation \cite{Das:2014fea, Arhrib:2012ia, Kanemura:1993hm}. The unitarity of the S-matrix demands the absolute eigenvalues of the scattering matrix should be less than $8\pi$ up to a particular scale. In our potential, bounds come from the eigenvalues of the corresponding S-matrix are as follows,
%%%%%%%%%%%%%%%%%%%%%%%%%%%%%%%%%%%%%%%%%%%%%%%%%
%%%%%%%%%%%%%%%%%%%%%%%%%%%%%%%%%%%%%%%%%%%%%%%%%
%%%%%%%%%%%%%%%%%%%%%%%%%%%%%%%%%%%%%%%%%%%%%%%%%
%%%%%%%%%%%%%%%%%%%%%%%%%%%%%%%%%%%%%%%%%%%%%%%%%
	\begin{equation}
			\begin{split}
				& |\lambda_3\pm \lambda_4 |\leq8\pi,
				~~~~ |\lambda_3\pm \lambda_5 |\leq 8\pi,\\
				&|\lambda_3+2\lambda_4\pm 3\lambda_5|\leq 8\pi,\\
				&\Big|\lambda_1+\lambda_2\pm \sqrt{(\lambda_1-\lambda_2)^2+\lambda_4} \Big|\leq 8\pi,\\
				&\Big|3\lambda_1+3\lambda_2\pm \sqrt{9(\lambda_1-\lambda_2)^2+(2\lambda_3+\lambda_4)^2}\Big|\leq 8\pi,\\
				&\Big| \lambda_1+\lambda_2\pm \sqrt{(\lambda_1-\lambda_2)^2+\lambda_5} \Big|\leq 8\pi.
			\end{split}
	\end{equation}

 \section{Baryogenesis in scotogenic model}\label{eq:8}
A fascinating way to dynamically produce the observed baryon asymmetry of the Universe (BAU) is via the mechanism of leptogenesis \cite{leptogenesis}. There arises an intrinsic limitation of the standard thermal leptogenesis, which is due to the requirement of a very high right-handed neutrino (RHN) mass scale. In the most generic scenario, occasionally known as the vanilla leptogenesis, there exists an absolute lower bound on the mass of the lightest RHN to be $M_{1} \simeq 10^{9}$ GeV \cite{Davidson:2002qv,Buchmuller:2002rq}. Whereas, in the case of the scotogenic model, with three $Z_{2}$ odd SM singlet fermions, one can bring down the limit on the lightest RHN mass scale to be as low as 10 TeV \cite{Hugle:2018qbw,Borah:2018rca}. In our work, we have taken the lightest RHN mass scale of the range $10^{4}-5\times10^{5}$ GeV, and that of the heavier RHNs, $N_{2}$ and $N_{3}$ of the range $10^{7}-5\times10^{8}$ GeV and $10^{12}-5\times10^{13}$ GeV respectively for generating the required baryogenesis. Since it is kinematically allowed via the Yukawa interactions, the SM singlet neutral fermions decay into the SM leptons, and the inert Higgs doublet $\eta$. In our work, we have considered the non-thermal production of DM within the IHDM desert via late decays of $N_{1}$ and thereby a small decay width of $N_{1}$ is considered for it to decay after DM freezes-out. Again, due to the consideration of the IHDM desert, the freeze-out will occur below the sphaleron temperature. Thus, the lifetime of $N_{1}$ will be more than sphaleron time prohibiting its decay to lepton asymmetry in to the observed baryon asymmetry above the sphaleron scale. Hence, lepton asymmetry is generated only because of the asymmetry created by the decay of $N_{2}$, which is the next to lightest RHN. The asymmetry produced by $N_{3}$ decays is considered negligible as a result of strong washout effect mediated by $N_{2}$ or $N_{3}$ itself. This leptogenesis is further converted into the baryon asymmetry of the Universe (BAU) by the electro-weak sphaleron phase transition \cite{Dine:2003ax}. The simultaneous Boltzmann equations for $N_{2}$ decay and formation of $N_{B-L}$ are to be solved to obtain the results for baryogenesis. The B-L calculation is mainly governed on the comparison between the Hubble parameter and the decay rates for $N_{2}\rightarrow l\eta, \bar{l}\eta^{*}$ processes, which will have a certain impact on the asymmetry as well as on the CP-asymmetry parameter $\epsilon_{2}$. We now further look into the various expressions and quantities that are required for the calculation of thermal leptogenesis in the scotogenic model. As essential in thermal leptogenesis, we need to distinguish between a weak washout and a strong washout regime. The differentiation is characterized based on the values of the decay parameter,
\begin{equation}
K_{2}= \frac{\Gamma_{2}}{H(z=1)},
\end{equation}
where, $\Gamma_{2}$ is the total $N_{2}$ decay width, $H$ being the Hubble parameter and $z= \frac{M_{2}}{T}$ with temperature $T$ of the photon bath.
Leptogenesis occurs above the electroweak scale during the era of radiation domination. The Hubble parameter can therefore be expressed in terms of $T$ as follows: 
\begin{equation}
H = \sqrt\frac{8\pi^{3}g_{*}}{90}\dfrac{T^{2}}{M_{Pl}},
\end{equation}
where $g_{*}$ is the effective number of relativistic degrees of freedom and $M_{Pl}\simeq 1.22\times 10^{19}$ GeV is the Planck mass.
With the varied choice of parameters, i.e., $M_{2}$, $M_{DM}$ and most crucially value of the lightest active neutrino mass, $m_{l}=10^{-13}$ eV compels the 3RHN scenario to fall in the strong washout regime similar to 2RHN case or type-I leptogenesis\cite{Hugle:2018qbw}. In the 2RHN case, only two active neutrinos are massive and the distinction between normal hierarchy(NH) and inverted hierarchy(IH) is made. However, for 3RHN the masses of the heaviest and the lightest active neutrino is almost same which results in the disappearence of the distinction between NH and IH. Also, the 2RHN always falls in the strong washout regime as the decay parameter ($K_2$) has values greater than $10^{3}$ for larger parameter space. % Thus we get a significantly large value of $K_{1}$, which is $K_{1}\simeq 10^{3}$ and above. This further enables us to assume $N_{1}$ dominated leptogenesis and neglect washout via scattering effects. Thereby, for a large value of $K_{1}$, we can use the approximation for the efficiency factor in the strong washout regime as,
%\begin{equation}
%\kappa_{1}(K_{1})\simeq \frac{1}{1.2K_{1}[\text{ln}K_{1}]^{0.8}}.
%\end{equation}
 The $N_{2}$ decay rate incorporating the Yukawa coupling is given by,
\begin{equation}
\Gamma_{2}= \frac{M_{2}}{8\pi}(Y^{\dagger}Y)_{22}\left[1- \Big(\frac{m_{DM}}{M_{2}}\Big)^{2}\right]^{2}= \frac{M_{2}}{8\pi}(Y^{\dagger}Y)_{22}(1-\eta_{2})^{2}.
\end{equation}
The CP asymmetry parameter $\epsilon_{2}$ for the decays $N_{2}\rightarrow l\eta, \bar{l}\eta^{*}$ is given by, 
\begin{equation}
\epsilon_{2}= \frac{1}{8\pi(Y^{\dagger}Y)_{22}}\sum_{j\ne 2}Im[(Y^{\dagger}Y)^{2}]_{2j}\left[ f(r_{j2},\eta_{2})- \frac{\sqrt{r_{j2}}}{r_{j2}-1}(1-\eta_{2})^{2}\right],
\end{equation}
where, the term $ f(r_{j2},\eta_{2})$ is expressed as,
\begin{equation}
f(r_{j2},\eta_{2})= \sqrt{r_{j2}}\left[1+ \frac{(1-2\eta_{2}+r_{j2})}{(1-\eta_{2})^{2}}  ln(\frac{r_{j2}-\eta_{2}^{2}}{1-2\eta_{2}+r_{j2}})\right],
\end{equation}
with $r_{j2}= \big(\frac{M_{j}}{M_{2}}\big)^{2}$, $\eta_{2}\equiv \big(\frac{m_{DM}}{M_{2}}\big)^{2}$.
The frequently appearing $Y^{\dagger}Y$ in the above equations can be expressed using the CI-parametrization\cite{Casas:2001sr},
\begin{equation}
(Y^{\dagger}Y)_{ij}= \sqrt{\Lambda_{i}\Lambda_{j}}(RM_{\nu}^{diag}R^{\dagger})_{ij}.
\end{equation}
An exciting piece of information regarding the $Y^{\dagger}Y$ is that it is independent of the PMNS matrix. This ensures that the CP-violating phases applicable for leptogenesis is independent of the CP-violating phases in PMNS matrix. In our work, we obtain the yukawa coupling matrix in the range $10^{-6}-1$. 
Again, starting with the initial thermal abundance of $N_{2}$, wherein its rate of interaction is above the Hubble rate, we solve the Boltzmann equations. It is only feasible if the Yukawa couplings corresponding to $N_{2}$ are not very small. In our work, we calculate the Yukawa coupling, which falls in the range applicable to generate the observed baryon asymmetry.

The Boltzmann equations for the number densities of $N_{2}$ and $N_{B-L}$, given by \cite{Davidson:2002qv},
\begin{equation}\label{eq:3}
\frac{dn_{N_{2}}}{dz}= -D_{2}(n_{N_{2}} - n_{N_{2}}^{eq}),
\end{equation}
\begin{equation}\label{eq:4}
\frac{dn_{B-L}}{dz}= -\epsilon_{2}D_{2}(n_{N_{2}} - n_{N_{2}}^{eq})- W_{2}n_{B-L},
\end{equation}
respectively. The equilibrium number density of $N_{2}$ is given by $n_{N_{2}}^{eq}= \frac{z^{2}}{2}K_{2}(z)$ , where $K_{i}(z)$ is the modified  Bessel function of $i^{th}$ type and
\begin{equation}
D_{2}\equiv \frac{\Gamma_{2}}{Hz} = K_{N_{2}}z\frac{K_{1}(z)}{K_{2}(z)}
\end{equation} 
is the measure of the total decay rate with respect to the Hubble rate, and $W_{2}$ is the total washout rate given by $ W_{2}= \frac{\Gamma_{W}}{Hz}$. The total washout term $W_{2}$ is the sum of the washout due to inverse decays $l\eta,\bar{l}\eta^{*}\rightarrow N_{1}$ and the washout due to the $\Delta L= 2$ scatterings $l\eta \leftrightarrow \bar{l}\eta^{*},ll \leftrightarrow \eta^{*}\eta^{*}$, i.e. $W_{2}= W_{2D}+W_{\Delta L=2}$ \cite{Hugle:2018qbw}, where $W_{2D}= \frac{1}{4}K_{N_{2}}z^{3}K_{1}(z)$
and,
\begin{equation}\label{eq:1}
W_{\Delta L=2} \simeq \dfrac{18\sqrt{10}M_{Pl}}{\pi^{4}g_{l}\sqrt{g_{*}}z^{2}v^{4}}(\frac{2\pi^{2}}{\lambda_{5}})^{2}M_{2}\bar{m_{\varsigma}}^{2}.
\end{equation}
In Eq.\eqref{eq:1}, $g_{l}$ stands for the internal degrees of freedom for the SM leptons, and $\bar{m_{\varsigma}}$ is the effective neutrino mass parameter, defined by:
\begin{equation}
\bar{m_{\varsigma}}^{2} \simeq 4\varsigma_{1}^{2}m_{1}^{2} + \varsigma_{2}m^{2_{2}} +\varsigma_{3}^{2}m_{3}^{2},
\end{equation}
with $m_{i}'s$ being the light neutrino mass eigenvalues and $ \varsigma_{k}$ is as defined as:
\begin{equation}
\varsigma_{k} = \Big(\frac{M_{k}^{2}}{8(m_{\eta_{R}^{0}}^{2}-m_{\eta_{I}^{0}}^{2})}[L_{k}(m^{2}_{\eta_{R}^{0}}) - L_{k}(m^{2}_{\eta_{I}^{0}})]\Big)^{-1}.	
\end{equation} 
We assess the final B-L asymmetry $n_{B-L}^{f}$ just before sphaleron freeze-out by numerically solving the Eqs.\eqref{eq:3} and \eqref{eq:4}, which is further converted into the baryon-to-photon ratio as,
\begin{equation}
n_{B}= \frac{3}{4}\frac{g_{*}^{0}}{g_{*}}a_{sph}n_{B-L}^{f}\simeq 9.2\times 10^{-3}n_{B-L}^{f},
\end{equation} 
where $a_{sph}=\frac{8}{23}$ is the sphaleron conversion factor with the consideration of two Higgs doublet. $g_{*}= 110.75$ is the effective relativistic degrees of freedom at the time of final lepton asymmetry production, and $g_{*}^{0}= \frac{43}{11}$ is the effective degrees of freedom at the recombination epoch. In this work, we have studied the effects on leptogenesis by the variation of parameters such as quartic coupling in the range $ 10^{-5}- 5$, the probable DM candidate mass in the intermediate-mass regime, i.e., $M_{W} < M_{DM} \le550$ GeV. From this choice of parameters, along with the mass of the lightest neutrino mass in the range $10^{-13}$ eV for both NH and IH, we calculate the Yukawa couplings for which we achieve $n_{B}^{obs}$ inferred from the Planck limit 2018, i.e., $(6.04\pm0.08)\times 10^{-10}$ at $68\%$ C.L. \cite{Aghanim:2018eyx}. Therefore, we get baryogenesis keeping intact the light neutrino mass satisfying the neutrino oscillation data. 
\section{Neutrinoless double beta decay}\label{eq:NDBD}
With the light neutrino parameters considered in our work, we can make connections with observable in the on-going experiments. A well known and significant experimental technique of detecting neutrino mass is the neutrinoless double beta decay ($0\nu\beta\beta$) \cite{Mohapatra:1986su,Barry:2013xxa,Borgohain:2017inp}, with experiments such as KamLAND-Zen, GERDA, KATRIN. In such experiments, what measured is the effective neutrino mass $|m_{\beta\beta}|$ which can be determined by the formula,
\begin{equation}
|m_{\beta\beta}|= \sum_{k=1}^3m_{k}U_{ek}^{2}\label{eq:12}
\end{equation}
where, $U_{ek}^{2}$ are the elements of the PMNS matrix with $k$ holding up the generation index. This eq.\eqref{eq:12} can be further expressed as,
\begin{equation}
|m_{\beta\beta}|= |c_{12}^{2}c_{13}^{2}m_{1} + s_{12}^{2}c_{13}^{2}m_{2}e^{2i\alpha} + s_{13}^{2}m_{3}e^{2i\beta}|
\end{equation}
where, $c_{ij}$= $\cos\theta_{ij}$ and $s_{ij}$= $\sin\theta_{ij}$. It is important to check the satisfying bound of the effective mass with the lightest neutrino mass so that we can relate the current light neutrino parameters giving correct hints to ongoing experiments and their future sensitivity.
\section{Dark matter in scotogenic model}\label{eq:7}
The dark matter, which was in chemical and thermal equilibrium in the early Universe, loses its equilibrium state when the pair annihilation rate becomes less than the expansion rate of the Universe, eventually leading the particles to decouple from the cosmic plasma.
The relic densities of such thermally produced dark matter candidates can be calculated by solving the Boltzmann equation \cite{Scherrer:1985zt,Kolb:1990vq}:
\begin{equation}
	\dot{n}_{DM} + 3Hn_{DM} = -<\sigma v> (n^{2}_{DM}- (n^{eq}_{DM})^{2}),
\end{equation}
where, $n_{DM}$ is the number density of the dark matter candidate and $n_{DM}^{eq}$ is the number density of the dark matter candidate in thermal equilibrium. The numerical solution of the Boltzmann equation in terms of partial wave expansion, $<\sigma v> = a + bv^{2}$ is of the form,
\begin{equation}
	\Omega h^{2}\approx \frac{1.04 \times 10^{9} x_{f}}{M_{Pl}\sqrt{g_{*}}(a + 3b/x_{f})},
\end{equation}
where, $x_{f}= \frac{m_{DM}}{T_{f}}$, $T_{f}$ is the freeze-out temperature, also  $v^{2}\simeq\frac{6}{x_{f}}$, $m_{DM}$ is the mass of dark matter, $g_{*}$ is the number of relativistic degrees of freedom at the time of freeze-out, and $M_{Pl} \approx 1.22 \times 10^{19}$ GeV is the Planck mass. Furthermore, we can also express this above expression in a simpler analytical form for the approximation of DM relic abundance as \cite{Jungman:1995df},
\begin{equation}
	\Omega h^{2} \approx \dfrac{3 \times 10^{-27} cm^{3} s^{-1}}{<\sigma v>}
\end{equation}
The corresponding thermal averaged annihilation cross section is therefore given by\cite{Gondolo:1990dk};
\begin{equation}
	<\sigma v> = \dfrac{1}{8m_{DM}^{4}TK_{2}^{2}(m_{DM}/T)}\int_{4m_{DM}^{2}}^{\infty}\sigma (s-4m_{DM}^{2})\sqrt{s}K_{1}(\sqrt{s}/T) ds ,
	\end{equation}
where, $K_{1}$ and $K_{2}$ are the modified Bessel functions, $m_{DM}$ is the mass of dark matter candidate and $T$ is the temperature.
In our model, we have considered one of the neutral component of the scalar doublet $\eta$ , i.e,  $\eta^{0}$ to be the dark matter candidate which resembles that with the inert doublet model discussed in the papers \cite{LopezHonorez:2006gr,Ahriche:2017iar,Deshpande:1977rw,Cirelli:2005uq,Barbieri:2006dq,Ma:2006wm,Hambye:2009pw,Dolle:2009fn,Honorez:2010re,Gustafsson:2012aj,Borah:2017dfn,Goudelis:2013uca,Arhrib:2013ela,Bhattacharya:2019fgs,Borah:2019aeq}. 
%{\bf The low mass regime, wherein the DM mass $m_{\eta^{0}}\le M_{W}$, DM annihilation into the SM fermions through $s$-channel Higgs mediation is dominant. As mentioned in the literature \cite{IHDM8}, the possible DM annihilations $\eta^{0}\eta^{0} \rightarrow WW^{*}\rightarrow Wf\bar{f^{/}}$ can also be significant in the low mass regime. Also, the co-annihilations of $\eta^{0}_{R}, \eta^{\pm}~ \text {and}~ \eta^{0}_{I}$ subjective to the mass differences $m_{\eta^{\pm}}-m_{\eta^{0}_{R}}\equiv \Delta M_{\eta^{\pm}}, m_{\eta^{0}_{I}}- m_{\eta^{0}} \equiv \Delta M_{\eta^{0}_{I}} $ can also play a role in the generation of relic abundance of DM. A comprehensive study on this type of co-annihilations are being discussed in \cite{relic2,relic3,relic4}.}
 From the literature \cite{Griest:1990kh}, we can express the effective cross-section as,
\begin{equation}
	\sigma_{eff} = \sum_{i,j}^{N} <\sigma_{ij} v> \frac{g_{i} g_{j}}{g_{eff}^{2}}(1 + \Delta_{i})^{3/2}(1 + \Delta_{j})^{3/2} e^{(-x_{f}(\Delta_{i} + \Delta_{j}))},
\end{equation}
with, $\Delta_{i}= \frac{m_{i}- m_{DM}}{m_{DM}}$ and $
g_{eff}= \sum_{i=1}^{N} g_{i}(1+\Delta_{i})^{3/2} e^{-x_{f}\Delta_{i}}.$

In the above equation, $ m_{i}$ denotes the mass of the heavier inert Higgs doublet. Therefore, the expression for the thermally averaged cross section is given by 
\begin{equation}
	<\sigma_{ij} v> = \dfrac{x_{f}}{8 m_{i}^{2} m_{j}^{2} m_{DM} K_{2}(\frac{m_{i}x_{f}}{m_{DM}}) K_{2} (\frac{m_{j} x_{f}}{m_{DM}})}\times \int_{(m_{i}+ m_{j})^{2}}^{\infty}  \sigma_{ij}(s- 2(m_{i}^{2}+ m_{j}^{2})) \sqrt{s} K_{1} \big(\frac{\sqrt{s}x_{f}}{m_{DM}}\big)ds.
\end{equation}

%{\bf In our work, we have shown the relic abundance for some particular value of DM mass within the intermediate mass range of dark matter, i.e $M_{W} < M_{DM} \le 550$ GeV by the usage of a computational package {\tt MicrOmega 5.0.4}\cite{micromega}.} 
%{Due to the choice of RHN masses heavier than the DM mass, its influence in the dark matter sector is negligible, i.e. it doesn't alter the relic abundance generated for the lightest inert scalar.}\\
 The only parameters mainly affecting the relic is the DM-Higgs coupling ($\lambda_{L}$) and the mass differences between the inert scalars. By appropriate choice of $\lambda_{L}$ and mass splitting, it is possible to generate the correct relic abundance for DM mass around 500GeV.  However, it is impossible to get the observed relic density below 500 GeV of dark matter mass, if the dark matter is produced thermally. Hence, we approach the non-thermal production of dark matter production mechanisms and study its consequences within the IHDM desert.
%{\bf As we can see from fig. \ref{dm1} that incase of $\Delta M_{\eta^{\pm}}$ = $\Delta M_{\eta^{0}_{I}} $= 1 GeV, we do get correct relic for $M_{DM}\sim 530$ GeV, but for values below it, we are not successful in generating the relic abundance. Also for larger mass splitting, say if we consider, $\Delta M_{\eta^{\pm}}$ = $\Delta M_{\eta^{0}_{I}} = 10$ GeV, we don't obtain the relic as expected from the LEP constraints \cite{Gustafsson2007,Lundstrm2009} which roughly rules out $M_{\eta^{0}_{I}}-M_{\eta^{0}_{R}}> 8$ GeV. Therefore, these shortcomings lead us to proceed with the non-thermal production of relic abundance.}\\

A non-thermal contribution in the production of relic abundance can be useful in generating the correct relic for masses of dark matter within the IHDM desert. The addition of the non-thermal part can enhance the under-abundant relic, which was observed in the IHDM desert to satisfy the Planck limit. This can be actually achieved by the late decay of the heavy particle, in our case $N_{1}$ decays to DM and SM leptons, i.e. $N_{1}\rightarrow l\eta, \bar{l}\eta^{*}$, resulting in the production of a correct relic of the DM candidate($\eta^{0}_{R}$).
% As $N_{1}$ decays out of equilibrium, it can also lead to successful leptogenesis. In the first part of our work, we encounter the underabundance of DM via thermal production, thereby, an another approach of introducing the non-thermal production along with the thermal production is brought into the scenario, in order to bring the underabundant DM to the observed value.
 We proceed with the method as discussed in \cite{Drees:2006vh}, and solve the coupled Boltzman equations shown below to calculate the number densities of DM candidate and $N_{1}$:
\begin{equation}
\begin{split}
	\frac{dn_{DM}}{dt}+3Hn_{DM}=& -<\sigma v>(n_{DM}^{2}-(n_{DM}^{eq})^{2})+N\Gamma_{N_{1}}n_{N_{1}},\\
	\frac{dn_{N_{1}}}{dt}+3Hn_{N_{1}}=& -\Gamma_{N_{1}}n_{N_{1}},
	\end{split}
\end{equation}
where $N$ is the average number of DM particles produced on the decay of $N_{1}$, and $\Gamma_{N_{1}}$ is the decay width of $N_{1}$.  We then move towards the analytical solution of the Boltzmann equation for $n_{N_{1}}$ by taking into consideration some of the crucial assumptions, that the co-moving entropy density($g_{*s}$) and co-moving energy density($g_{*}$) is almost constant. We now transform the above equation interms of $Y_{DM}$ and $Y_{N_{1}}$ by using the relation $Y_{DM}= \frac{n_{DM}}{s}$ and $Y_{N_{1}}= \frac{n_{N_{1}}}{s}$ where $s= \frac{2\pi^{2}g_{*s}T^{3}}{45}$ is the entropy density. The final equation we obtain on changing the variable t to $x=\frac{M_{DM}}{T}$ and also inserting the above variables:
\begin{equation}\label{eq21}
\frac{dY_{DM}}{dx}= -\frac{<\sigma v>s}{Hx}(Y_{DM}^{2}-(Y_{DM}^{eq})^{2})+NrxY_{N_{1}}(x_{0})exp(-\frac{r}{2}(x^{2}-x_{0}^{2})).
\end{equation}  
In eq.\eqref{eq21}, $r= \frac{\Gamma_{N_{1}}}{Hx^{2}}= \Big(\frac{\Gamma_{N_{1}}M_{Pl}}{\pi M_{DM}^{2}}\Big)\sqrt{\dfrac{90}{g_{*}}}$ is a constant depending upon the deacy width of the heavy decaying particle and $Y_{N_{1}}(x_{0})$ is the initial abundance of $N_{1}$. After finding the numerical solution of eq.\eqref{eq21}, we obtain the present day abundance of DM and further we implement this solution in calculating the relic abundance of DM in the present Universe using the equation:
\begin{equation}
\Omega h^{2}= \frac{M_{DM}Y_{0}s_{0}}{\rho_{c}},
\end{equation}
where, $\rho_{c}\sim 1.05\times 10^{-5}h^{2}$ GeV$cm^{-3}$ is the critical density of the Universe, $s_{0}\sim 2891.2~ cm^{-3}$ is the current entropy density and $h=0.72$ is the Hubble parameter. 

As we know, the decay of $N_{1}$ release entropy which may mimic the abundance light element that are involved in the big-bang nucleosynthesis (BBN). Hence, the decay of $N_{1}$ must not occur during or after the epoch of the BBN  \cite{Borah:2017dfn}. Thus, we get a constraint on the minimum value of decay width of $N_{1}$, i.e., $\Gamma_{N_{1}}\geq \Gamma_{N_{1},min}\equiv 6.58\times 10^{-25}$ GeV, arising from the consideration that the decay lifetime of $N_{1}$ should be less than 1 second. Again, an upper bound on the decay width, i.e. $\Gamma_{N_{1}}\leq \Gamma_{N_{1},max}\equiv \frac{M_{DM}^{2}}{x_{0}}\times 10^{-18}$ GeV is a manifestation of the fact that the decay of $N_{1}$ should take part mostly after the DM candidate freezes out thermally so as to give adequate contribution towards the relic abundance. Thus, we investigate the limitations that we encountered during the thermal production of the relic and see for what benchmark values of the free parameters $\Gamma_{N_{1}}$ and $Y_{N_{1}}(x_{0})$ we can have correct relic abundance within the IHDM desert even for high mass splitting.

%We also scan the parameter space for $\lambda_{L}- M_{DM}$ satisfying the relic abundance and direct detection constraint from XENON1T, thereby, we proceed with spin independent scattering cross section calculation. This type of cross section can be a product of the scalar dark matter in this work. 
As we have considered the lightest stable scalar particle to be a probable dark matter candidate, thus, the spin independent scattering cross section of the SM Higgs is expressed by\cite{Barbieri:2006dq}:
\begin{equation}
\sigma_{SI}= \frac{\lambda_{L}^{2}f^{2}m_{\mu}^{2}m_{n}^{2}}{4\pi m_{h}^{4}M_{DM}^{2}}
\end{equation} 
where, $\lambda_{L}= (\lambda_{3}+\lambda_{4}+\lambda_{5})/2$ is the quartic coupling taking part in the DM-Higgs interaction, $m_{\mu}^{2}= m_{n}M_{DM}/m_{n}+M_{DM}$ is the DM-nucleon reduced mass and $f$ is the Higgs-nucleon coupling which is estimated to be $f=0.32$ \cite{Giedt:2009mr}. There also can be a Higgs portal coupling independent DM-nucleon scattering cross-section at a one-loop level \cite{Klasen:2013btp}. However, by appropriate choice of the mass splitting between the scalar components, we can generate spin-independent scattering cross-section much lower than that obtained from direct detection experiment XENON1T. 
 \section{Numerical analysis and results}\label{eq:9}

 \begin{figure}
	\includegraphics[width=0.4\textwidth]{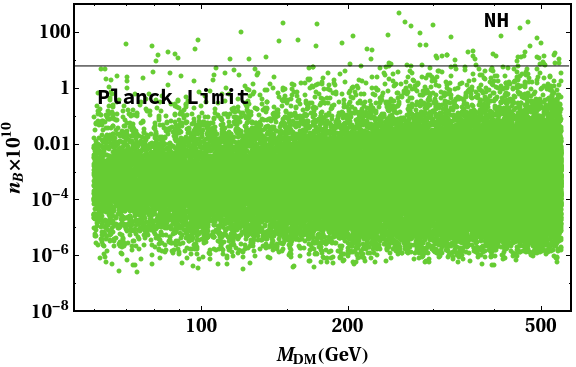}
	\includegraphics[width=0.4\textwidth]{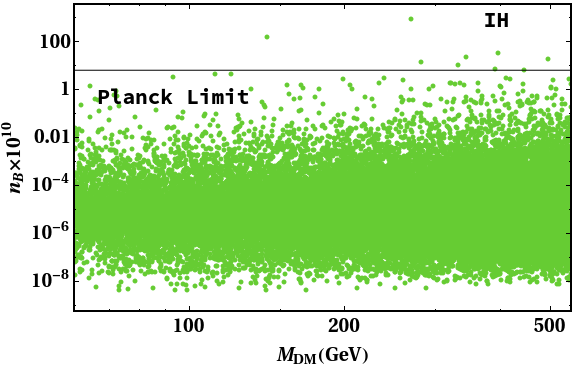}\\
	\includegraphics[width=0.4\textwidth]{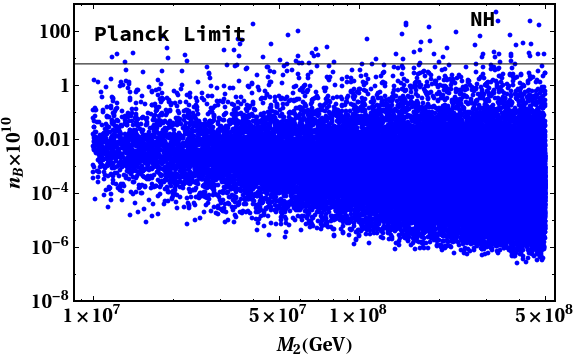}
	\includegraphics[width=0.4\textwidth]{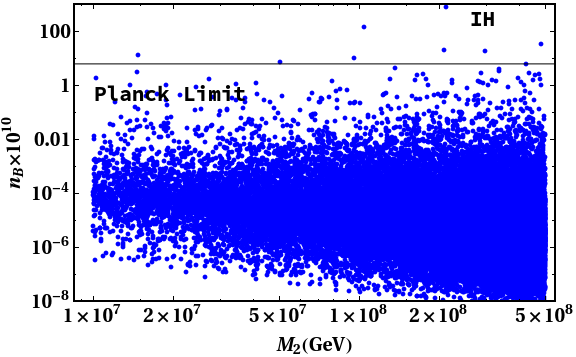}\\
	\includegraphics[width=0.4\textwidth]{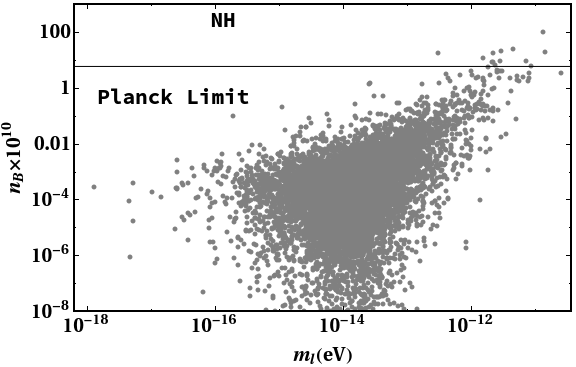}
	\includegraphics[width=0.4\textwidth]{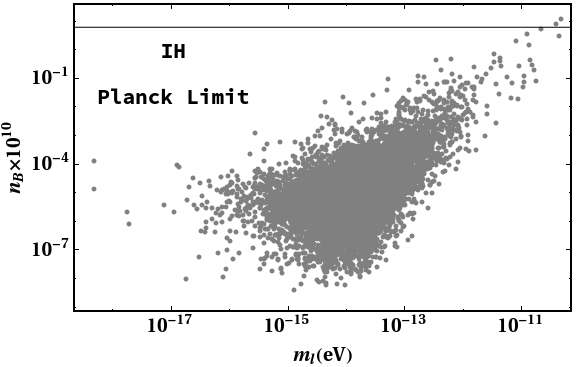}\\

	\includegraphics[width=0.4\textwidth]{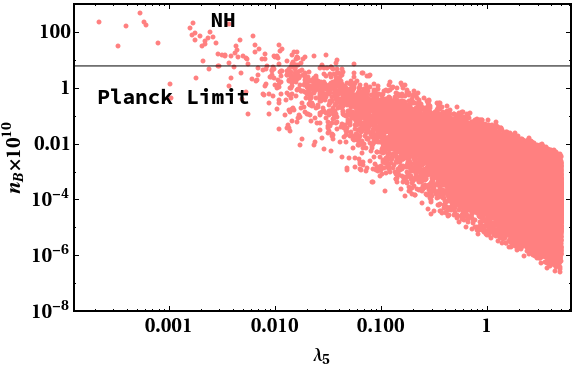}
	\includegraphics[width=0.4\textwidth]{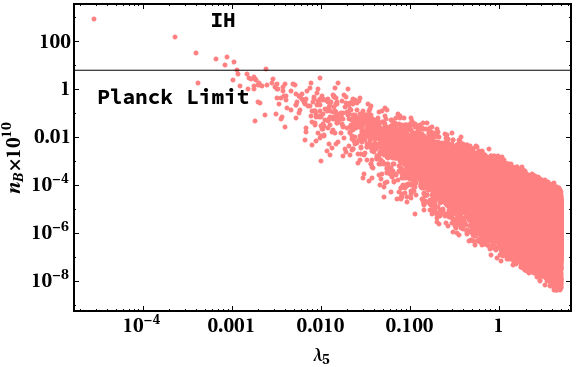}\\
	\caption{Plots in the first-row show baryon asymmetry as a function of dark matter mass ($M_{DM}$), the second-row show baryon asymmetry as a function of right-handed neutrino mass ($M_{2}$), in third-row baryon asymmetry as a function of the lightest neutrino mass eigenvalue is shown. The fourth row depicts baryon asymmetry as a function of the absolute value of quartic coupling ($|\lambda_5|$) for NH and IH, respectively. The black horizontal line gives the current Planck limit for BAU. }\label{bau1}
\end{figure}

\begin{figure}[h]
	\begin{center}
		\includegraphics[width=0.43\textwidth]{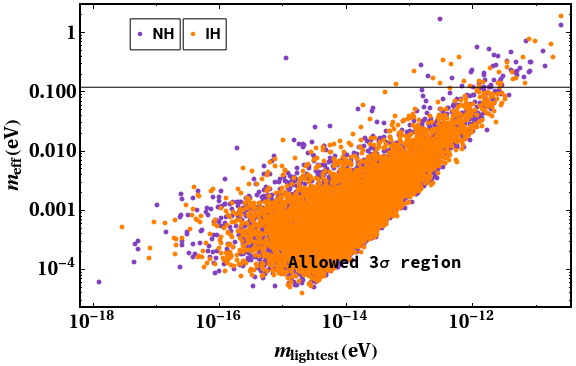}\\
	\caption{	 Effective mass as a function of lightest neutrino mass eigenvalue for NH/IH. The horizontal(black) line is the upper limit for the effective mass ($m_{\beta\beta}(eV)\sim0.12(eV)$) of light neutrinos obtained from KamLAND-Zen experiment and the red vertical line depicts the upper bound given by the Planck limit for the summation of light neutrino masses.}\label{eff1}
		
	\end{center}
\end{figure}
\begin{figure}[h]
	\begin{center}
		\includegraphics[width=0.43\textwidth]{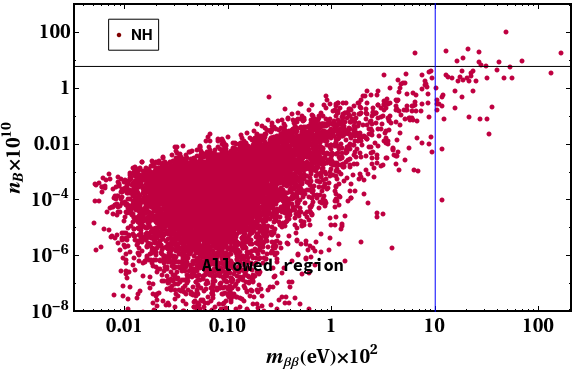}
		\includegraphics[width=0.43\textwidth]{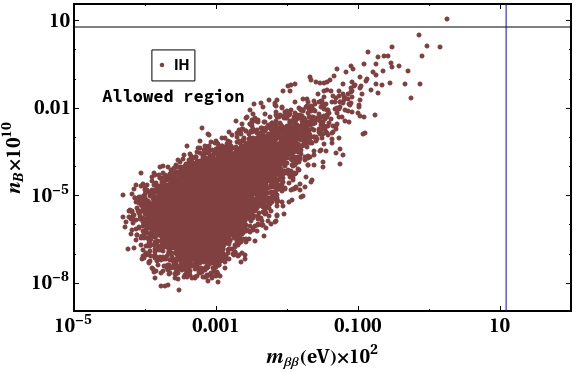}\\
		\caption{ Baryon asymmetry as a function of effective mass of neutrino for NH/IH. The horizontal(black) line is the Planck limit for BAU and the vertical(blue) line depicts the KamLAND-Zen limit for $0\nu\beta\beta$.}\label{bareff2}
	\end{center}
\end{figure}
In this study, we choose the dark matter mass in the intermediate-mass range, $M_{W} < M_{DM} \le 550$ GeV, and study the consequences of neutrino mass, neutrinoless double beta decay and baryon asymmetry of the Universe. The plot in the first row of fig. \ref{bau1} depicts that the observed baryogenesis is satisfied for almost the entire IHDM desert for NH, whereas, in case of IH, baryogenesis is obtained for dark mass above 300 GeV with very scanty points. Furthermore, for $N_{2}$ leptogenesis in the scotogenic model, we can conclude the mass of the next to the lightest RHN must be greater than $10^{7}$ GeV, which has risen up the TeV scale thereby enhancing the washout effect.
Hence, in our work, we have chosen the RHN masses $M_{1}$, $M_{2}$ and $M_{3}$ in the range $10^{4}-5\times10^{5}$ GeV, $10^{7}-5\times10^{8}$ GeV and $10^{12}-5\times10^{13}$ GeV respectively. 

The first row of fig. \ref{bau1} shows the variation between the baryon asymmetry of the Universe and the dark matter mass ($M_{DM}$). In the second row we have the variation of BAU results with the mass of the next to lightest RHN $M_{2}$ for both NH and IH and thus obtain the parameter space of $M_{DM}$ and $M_{2}$ that satisfies the currently observed value of BAU in both the mass orderings. From the results of $M_{2}$ vs. $\eta_{B}$, we see that the entire range chosen for $N_{2}$ generates BAU, whereas in the case of IH, very few points above $5\times 10^{7}$ GeV satisfies the Planck limit for BAU.
  \begin{figure}[h]
 	\begin{center}
 		\includegraphics[width=0.42\textwidth]{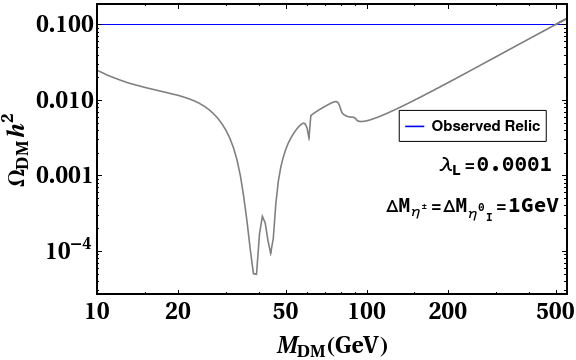}
 		\includegraphics[width=0.42\textwidth]{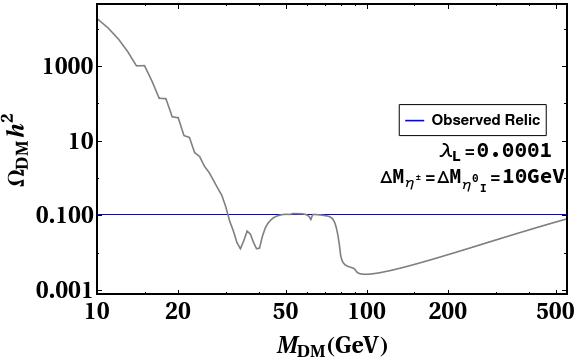}
 		\caption{Variation of relic abundance of DM in the intermediate dark matter mass range. Planck limit for the observed relic abundance is given by the horizontal(blue) line. The V shaped portion at around $M_{DM}\approx M_{h}/2$ is the resonance in the annihilation of DM into SM fermions mediated via Higgs boson(h) in the s-channel.}	\label{dm1}
 	\end{center}
 \end{figure}
 \begin{figure}[h]
 	\begin{center}
 		\includegraphics[width=0.42\textwidth]{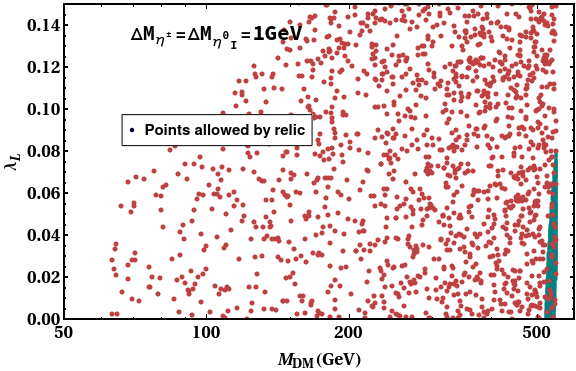}
 		\includegraphics[width=0.42\textwidth]{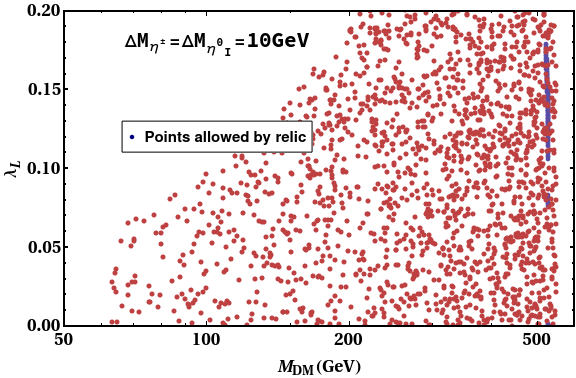}
 		\caption{ The allowed region of parameter space in $\lambda_{L}$-$M_{DM}$ plane from the requirement of satisfying the relic abundance and depiction of the strict constraints from dark matter direct detection experiment,XENON1T. The red points corresponds to the region allowed by direct detection experiment and the small vertical lines in both the panels of the figure are the points that generate the observed relic abundance.}\label{dm2}
 	\end{center}
 \end{figure}
%As we realize WIMP type of dark matter, we seek the Yukawa couplings to be much larger than that required in FIMP type dark matter. We can generate large values of Yukawa couplings by taking the lightest active neutrino mass($m_{l}$) in the range of $10^{-3}-1$ eV. Such a mass range of $m_{l}$ along with the fact that $M_{DM}\neq M_{N_{1}}$($\eta_{1}\neq 1$) is crucial in the strong washout regime. Thus, even for the 3RHN scenario, we are in a strong washout regime similar to the 2RHN case. 
% For lightest active neutrino mass choosen in the range, $m_{l}= 10^{-19}-10^{-13}$ eV,
We calculate the mass eigenvalues of light neutrinos for the scotogenic model by keeping some variables($M_{k}$,$\lambda_5$,$\eta_{R}^{0}$ and $\eta_{I}^{0}$) fixed as mentioned earlier and solving the model parameters. 
  A plot of baryogenesis vs. the lightest neutrino mass eigenvalue, $m_{l}$ is shown in the third row of fig.\ref{bau1}, where the left panel shows the variation for NH and the right panel for IH. Scanning the whole parameter space, we can clearly see that for NH, there are few points in the range $m_{l}= 10^{-13}-10^{-12}$ eV, satisfying the Planck limit for observed baryogenesis. However for IH, the points satisfying baryogenesis becomes very scarce. Thus, we can conclude that the NH is more preferable in terms of BAU than IH in our study. The entire work is carried out for quartic coupling $|\lambda_{5}|=10^{-5}-5$. Therefore, we analyze the parameter space of the quartic coupling satisfying the observed baryon asymmetry, which can be estimated to be $\mathcal{O}$($10^{-2}- 5$) as shown in the last row of fig. \ref{bau1} for NH. But the same analysis differ incase of IH, wherein very few points below $|\lambda_{5}|=10^{-2}$ satisfies BAU. Thus, the parameter space taken in our study is more inclined towards generating BAU for NH compared to IH. As we have also studied $0\nu\beta\beta$ in this work and the variation of $m_{\beta\beta}$ vs. $m_{l}$ for NH and IH are shown in fig.\ref{eff1}. Here, the horizontal line is the upper limit for the effective mass of active neutrinos obtained from KamLAND-Zen experiment. Thus, we can see that our study satisfies this constraint as maximum of the points for both NH and IH lie below the upper limit. Moreover, a correlative analysis of the points satisfying both effective mass and baryogenesis is also shown in fig.\ref{bareff2}. This draws an interesting result as we have seen points satisfying both BAU and $0\nu\beta\beta$ in NH. Whereas for IH, we merely have same points obeying BAU and $0\nu\beta\beta$ simultaneously. 

The probable candidate of DM will be the lightest particle among the inert Higgs doublet. In our study, $\eta_{R}^{0}$ is considered to be a source of DM, with the assumption of it being the lightest of all scalars. Therefore, it's relic abundance is calculated by implementing first this minimal scotogenic model in {\tt Feynrules}\cite{Alloul:2013bka} and then using the computational package \texttt{MicrOmega 5.0.4}\cite{Belanger:2018ccd}. The relic abundance as a function of the DM mass $M_{DM}$ is manifested in fig.\ref{dm1}, where, the DM-Higgs coupling is taken to be as low as $\lambda_{L}=0.0001$ and the mass differences $\Delta M_{\eta^{\pm}}= \Delta M_{\eta^{0}_{I}}= 1$ GeV (left panel). Also, in fig.\ref{dm1}, we have shown a similar plot of relic $vs. ~M_{DM}$ for higher values of $\Delta M_{\eta^{\pm}} = \Delta M_{\eta^{0}_{I}} = 10$ GeV (right panel). From fig.\ref{dm1}, we can anticipate that for low mass splitting between the scalars , i.e., $\Delta M_{\eta^{\pm}} = \Delta M_{\eta^{0}_{I}} = 1$ GeV, the relic is suppressed in the low mass regime due to the increase in co-annihilation between the different components of inert scalar doublet. Whereas, in the high mass regime for $\Delta M_{\eta^{\pm}} = \Delta M_{\eta^{0}_{I}} = 10$ GeV the relic is suppressed relic because the annihilation contribution of the electroweak bosons increases with the mass square differences among the inert scalars.

Furthermore, instead of fixing the DM-Higgs coupling, we show the allowed region of parameter space in the $\lambda_{L}-M_{DM}$ plane from the obligation of satisfying the correct relic abundance depicted in fig.\ref{dm2}. With the relic abundance bound on the $\lambda_{L}-M_{DM}$ plane, there also exist strict constraint from the dark matter direct detection experiment XENON1T. The scattered points in fig.\ref{dm2} corresponds to the values of $M_{DM}$ and $\lambda_{L}$, which are allowed from the direct detection bound of XENON1T and the small dark portion refer to the points allowed by the current value of relic density. Thus, we can see that there exists a coincidence of both points signifying the parameter space, which obeys constraints from both the cosmological aspects mentioned above. We see a significant difference in the parameter space of $\lambda_{L}$ $w.r.t.$ the mass difference of the scalars. Hence, we can confirm the choice of mass difference is of utmost importance in determining the relic abundance of dark matter when we donot introduce the non-thermal production of DM. \cite{Klasen:2013btp}.
  \begin{figure}
		\begin{center}
			\centering
		\begin{subfigure}[b]{0.42\textwidth}
				\includegraphics[scale=0.34]{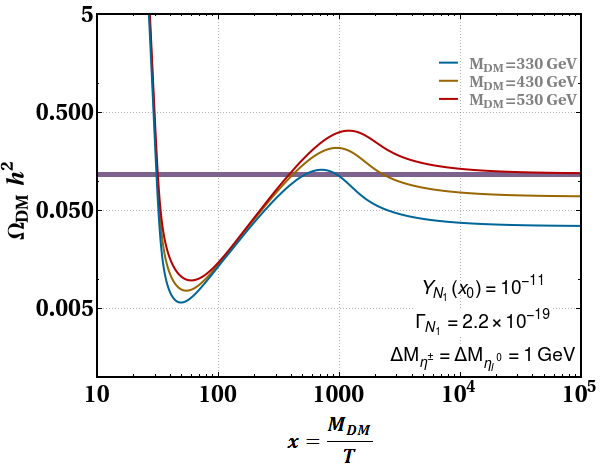}
			\caption{}
				\label{8a}
		\end{subfigure}
			\begin{subfigure}[b]{.42\textwidth}\centering
		\includegraphics[scale=0.34]{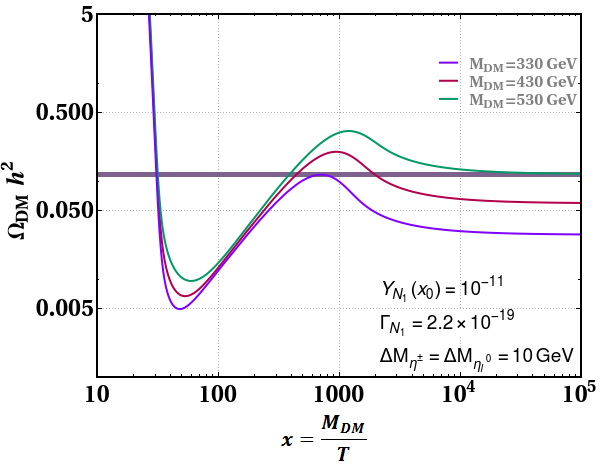}
		\caption{}
		\label{8b}
	\end{subfigure}
		\begin{subfigure}[b]{0.45\textwidth}\centering
	\includegraphics[scale=0.34]{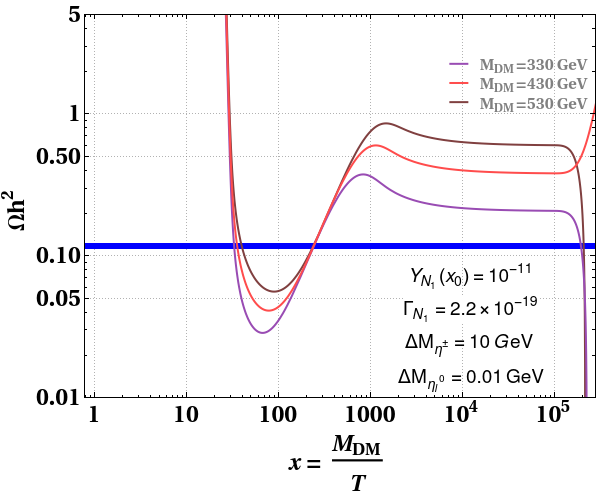}
\caption{}\label{8c}
\end{subfigure}
		%	\caption{ Variation of relic abundance for three different values of dark matter mass with fixed values of $\Gamma_{N_{1}}$ and $Y_{N}(x_{0})$. The corresponding parameters which contribute in determining relic are kept fixed with values: $\Delta M_{\eta^{\pm}}$ = $\Delta M_{\eta^{0}_{I}}= 10$ GeV, $\lambda_{L}= 0.0001$, $\lambda_{2}=0.2$ and $M_{h}= 125.5$ GeV.}
		\caption{ Variation of relic abundance vs $x(=M_{DM}/T$) for three different values of dark matter masses with fixed values of $\Gamma_{N_{1}}$ and $Y_{N_{1}}(x_{0})$ as given in the plot. The corresponding parameters which contribute in determining relic are kept fixed with values: (a) $\Delta M_{\eta^{\pm}}$ = $\Delta M_{\eta^{0}_{I}}= 1$ GeV, (b) $\Delta M_{\eta^{\pm}}$ = $\Delta M_{\eta^{0}_{I}}= 10$ GeV, and (c) $\Delta M_{\eta^{\pm}} =10$ GeV and $\Delta M_{\eta^{0}_{I}}= 0.01$ GeV, $\lambda_{L}= 0.0001$, $\lambda_{2}=0.2$ and $M_{h}= 125.5$ GeV.	}\label{dm8}
			
\end{center}
\end{figure}
\begin{figure}[ht]
		\begin{subfigure}{0.43\textwidth}
	\centering
		\includegraphics[width=\linewidth]{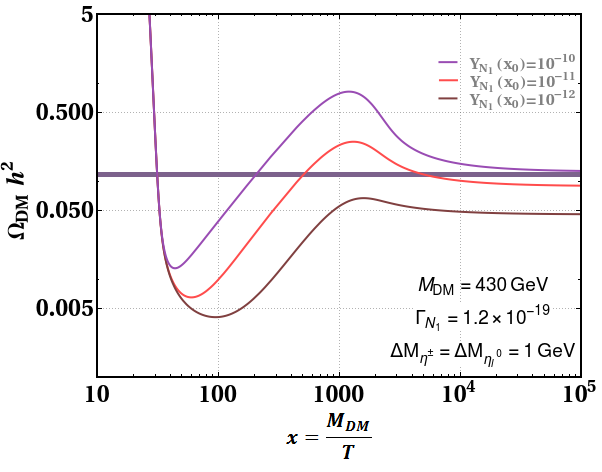}
			\caption{}
		\label{9a}
			\end{subfigure}
		\begin{subfigure}{0.43\textwidth}
			\centering
		\includegraphics[width=\linewidth]{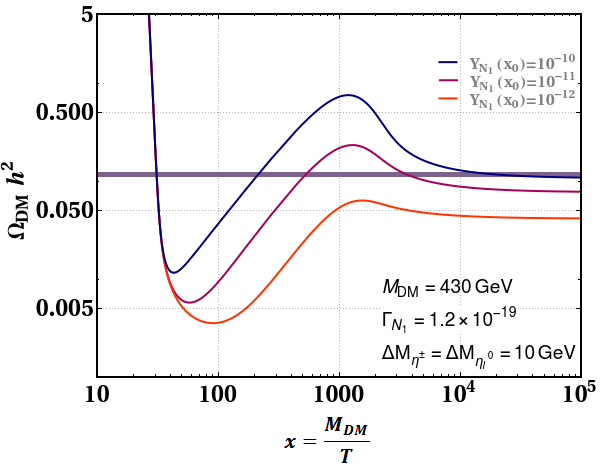}	
			\caption{}
		\label{9b}
	\end{subfigure}
	\begin{subfigure}{0.43\textwidth}
	\centering
			\includegraphics[width=.9\linewidth]{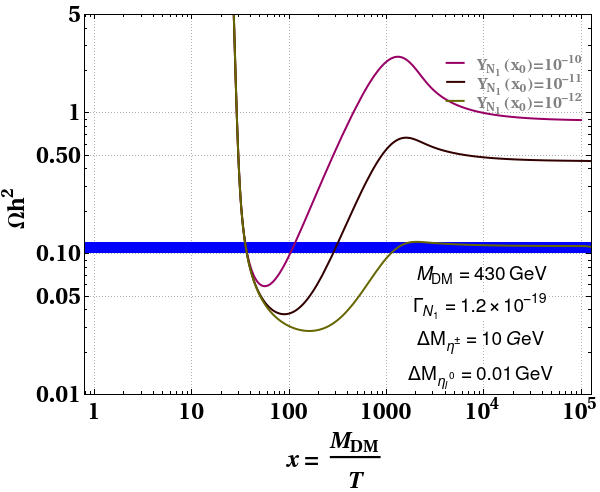}	
				\caption{}
			\label{9c}
				\end{subfigure}
		%	\caption{ Variation of relic abundance with DM mass fixed at $M_{DM}=430$ GeV and $\Gamma_{N}=2.5\times10^{-16}$ for three different values of $Y_{N}(x_{0}$. The corresponding parameters which contribute in determining relic are kept fixed with values: $\Delta M_{\eta^{\pm}}$ = $\Delta M_{\eta^{0}_{I}}= 10$ GeV, $\lambda_{L}= 0.0001$, $\lambda_{2}=0.2$ and $M_{h}= 125.5$ GeV. }
		
		\caption{ Plot of relic abundance vs $x(=M_{DM}/T$) with DM mass fixed at $M_{DM}=430$ GeV and $\Gamma_{N_{1}}= 1.2\times10^{-19}$ for three different values of  $Y_{N_{1}}(x_{0})$. The corresponding parameters which contribute in determining relic are kept fixed with values: (a) $\Delta M_{\eta^{\pm}}$ = $\Delta M_{\eta^{0}_{I}}= 1$ GeV, (b) $\Delta M_{\eta^{\pm}}$ = $\Delta M_{\eta^{0}_{I}}= 10$ GeV and (c) $\Delta M_{\eta^{\pm}} =10$ GeV and $\Delta M_{\eta^{0}_{I}}= 0.01$ GeV, $\lambda_{L}= 0.0001$, $\lambda_{2}=0.2$ and $M_{h}= 125.5$ GeV.}	\label{dm9}
	
\end{figure}

\begin{figure}[ht]
	\begin{center}
				\begin{subfigure}[b]{0.42\textwidth}
			\includegraphics[scale=0.35]{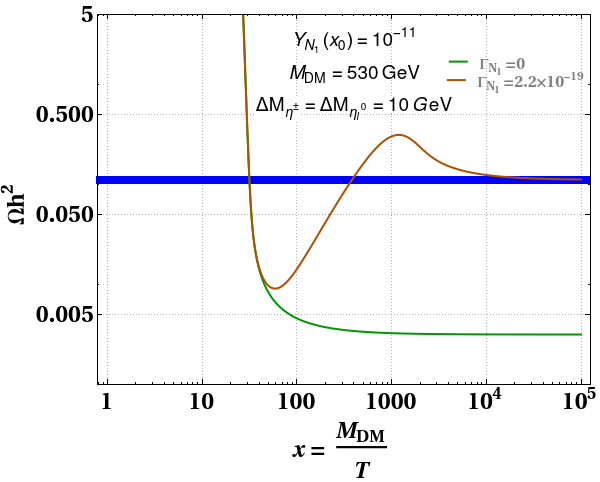}
			\caption{}
			\label{10a}
		\end{subfigure}
		\begin{subfigure}[b]{.42\textwidth}\centering
			\includegraphics[scale=0.35]{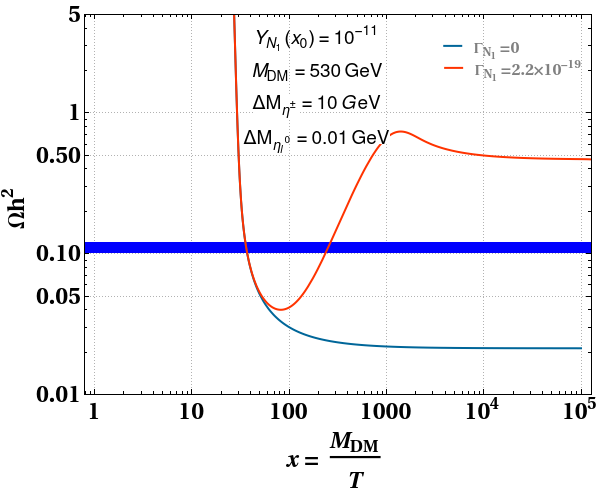}
			\caption{}
			\label{8b}
			\end{subfigure}		
		\caption{	 Relic abundance vs $x(=M_{DM}/T$) plot with dark matter mass fixed at 530 GeV and $Y_{N_{1}}(x_{0})=10^{-11}$ for two values of $\Gamma_{N_{1}}$. $\Gamma_{N_{1}}=0$ corresponds to thermal production and $\Gamma_{N_{1}}=2.2\times10^{-19}$ signifies non-thermal production. The corresponding parameters which contribute in determining relic are kept fixed with values: (a) $\Delta M_{\eta^{\pm}}$ = $\Delta M_{\eta^{0}_{I}}= 10$ GeV(left panel),  (b) $\Delta M_{\eta^{\pm}} =10$ GeV and $\Delta M_{\eta^{0}_{I}}= 0.01$ GeV(right panel), $\lambda_{L}= 0.0001$, $\lambda_{2}=0.2$ and $M_{h}= 125.5$ GeV.}\label{dm10}
	\end{center}
\end{figure}

Dark matter relic density primarily depends on the dark matter mass, Higgs portal coupling, and mass differences with the LSP and nLSP\footnote{Lightest stable particle and next to lightest stable particle.}. In the low mass region for $M_{DM}<10$ GeV, most dominating DM annihilation processes are to the SM fermions only, and due to small coupling strength and mass, we get an overabundance of the relic density. Moreover, the dominant part of the points ruled out by the Higgs/$Z$ invisible decay width and direct detection constraints for the low mass. Within IHDM, irrespective of the choice of parameter spaces, the region in between $M_W<M_{DM}\le530$ GeV does not give observed relic abundance value due to the very high annihilation rate of $DM + DM\rightarrow W^{\pm}W^{\pm}, ZZ$ \cite{Honorez:2010re, Khan:2015ipa, Das:2019ntw}. However, by considering different production mechanisms as discussed by \cite{Borah:2017dfn, Drees:2006vh}, we can work out the on the IHDM desert region to get observed relic abundance. Here we consider the decay of a particle $N_1$, which produces dark matter non-thermally, and by adjusting suitable decay width and initial abundance of dark matter candidate, we can generate observed relic density within the IHDM desert.  
From fig.\ref{dm1}, we can see the deviation in relic abundance, taking into consideration the crucial parameter, i.e., the mass splitting among the scalars of the inert doublet. For $\Delta M_{\eta^{\pm}}=\Delta M_{\eta^{0}_{I}}= 1$ GeV, we get the correct relic abundance corresponding to $M_{DM}= 530$ GeV, whereas for $\Delta M_{\eta^{\pm}}$ = $\Delta M_{\eta^{0}_{I}}= 10$ GeV, we fail to generate the relic. Therefore, we proceed with the non-thermal production of dark matter to see if the desired relic is obtained for the above value of dark matter and masses even lower than it. We consider the low mass splitting case with $M_{DM}$= 530 GeV and by appropriate choice of the decay width($\Gamma_{N_{1}}$), we see that for $Y_{N_{1}}(x_{0})= 10^{-11}$ GeV, it produces the correct relic abundance. Again for high mass splitting, the deviation in thermal and non-thermal production of the relic is observed. We verify the result obtained in the right panel of fig.\ref{dm1}, again by fig.\ref{dm10}, that dark matter is underabundant thermally. It shows that for $M_{DM}$=530 GeV and the choice of other parameters, $\Gamma_{N_{1}}= 2.5\times10^{-19}$ and $Y_{N_{1}}(x_{0})= 10^{-11}$, we obtain the relic, whereas for  $\Gamma_{N_{1}}=0$, there is an underabundant production of relic. We also do a relative study for three benchmark values of dark matter in the low mass splitting as well as the high mass splitting scenario depicted in fig.\ref{dm8}.
%(we may cut out this portion Fig.\ref{dm8} depicts the variation in relic for the different dark matter masses, which are further related to the annihilation cross-section($<\sigma v>$). As the dark matter mass decreases, $<\sigma v>$ increases, which results in the underabundance of the relic, as shown in fig.[12]. )
We now fine-tune the decay width in order to obtain the correct relic abundance for $M_{DM}=430$ GeV, which was underabundant for the values shown in fig.\ref{dm8}. Thus, fig.\ref{dm9} showcases the two different mass splitting scenarios for $M_{DM}=430$ GeV, and investigate the values of $\Gamma_{N_{1}}$ and $Y_{N_{1}}(x_{0})$ which satisfies the correct relic abundance.  %We again see the same pattern of variation in relic as was shown in fig.[14], but for different values of  $\Gamma_{N_{1}}$ and $Y_{N}(x_{0})$. As mentioned earlier, for  $M_{DM}$=530 GeV, we do get the correct relic taking in account the low mass difference case. Therefore, we try to probe the production of relic for smaller value of dark matter mass, in our case $M_{DM}$=430 GeV is taken as the benchmark value. From fig.[12], we can conclude the variation in relic is dependent on the choice of parameters, $\Gamma_{N_{1}}$ and $Y_{N}(x_{0})$.%
  We have also shown the variation of relic abundance for three benchmark values of dark matter in consideration with different values of scalar mass splittings, i.e $\Delta M_{\eta^{\pm}} =10$ GeV and $\Delta M_{\eta^{0}_{I}}= 0.01$ GeV in the fig. \ref{8c}. We see a deviation in the curves which were previously satisfying observed relic when similar mass splittings between the scalars were considered. However, for $M_{DM}=430$ GeV, $\Gamma_{N_{1}}= 1.2\times10^{-19}$ and $Y_{N_{1}}(x_{0})= 10^{-12}$ it is possible to generate the correct relic abundance inspite of the inequality between the values of scalar mass splittings. This can be seen in the fig. \ref{9c}. Also in the right panel of fig.\ref{dm10}, due to different values of scalar mass splittings as mentioned earlier, we see a vast deviation in the curves satisfying the relic abundance. 
Therefore, we can see a distinct variation of $\Delta M_{\eta^{\pm}}$, $\Delta M_{\eta^{0}_{I}}$, $\Gamma_{N_{1}}$ and $Y_{N_{1}}(x_{0})$ {\it w.r.t.} dark matter mass resulting in the production of correct relic abundance. 
\section{Conclusion}\label{eq:10}
In this paper, we study an extension of the SM popularly known as the {\it scotogenic model}, which is extended by a Higgs doublet ($\eta$) and three singlet neutral fermions ($N_{k}$). An additional $Z_2$ charge is assigned in the model, and all the SM particles are ever under it while additional fields are odd. The possibility of a DM candidate comes from the $Z_{2}$ odd lightest particle. We carry out this work with the dark matter mass strictly focusing in the intermediate dark matter mass range, also known as the inert Higgs doublet model (IHDM) desert, which lies between $M_{W} < M_{DM} \le 550$ GeV. Along with DM, baryogenesis via the mechanism of thermal leptogenesis and neutrinoless double beta decay is also addressed in this work. Leptogenesis is a result of the decay of $Z_{2}$ odd fermions, $i.e$, the heavy RHN, which occurs via the out-of-equilibrium decay into the SM leptons and the inert Higgs doublet. The out-of-equilibrium decay of $N_{2}\rightarrow l\eta, \bar{l}\eta^{*}$, where $\eta$ is the inert Higgs doublet constituting the dark matter candidate $\eta_{R}^{0}$ , generates the observed baryon asymmetry of the Universe. The final lepton asymmetry is generated only because of the asymmetry created by the decay of $N_{2}$, which is the next to lightest RHN. Again, for two different choice of mass splitting between the DM (LSP) and the next heavier scalar (nLSP), we study the relic abundance of the dark matter candidate (lightest of $\eta$). We also study the mixture of thermal and non-thermal production of DM abundance for various masses within the IHDM desert. In our study, the non-thermal DM within the IHDM desert is produced via late decays of $N_{1}$. Therefore, the lifetime of $N_{1}$ will be more than sphaleron time resulting in the discripency to generate the baryon asymmetry. This is because the decay width of $N_{1}$ considered in our work for the non-thermal production of DM is very small. Also we can say that the mass splitting between the inert scalars are crucial for thermal production of DM unlike that for non-thermal prodcution of DM. Although the inequality in the values of scalar mass splittings do create a difference in generating the observed relic abundance via non-thermal production.\\
%In this work we also focus on the  realization of the baryon asymmetry and dark matter in the scotogenic model keeping intact all the neutrino oscillation data and cosmological bounds. Meanwhile, we check if the out-of-equilibrium decay of $N_{1}\rightarrow l\eta, \bar{l}\eta^{*}$, where $\eta$ is the inert Higgs doublet constituting the dark matter candidate $\eta_{R}^{0}$ that can generate the observed baryon asymmetry of the Universe. Throughout the work, a fixed dark matter mass range, {\it i.e} the IHDM desert is considered.
% From literature it is known that no relic abundance is generated in this IHDM desert, however, in our work we have considered non-thermal production of DM and shown the relic abundance in this region by fine tunning of $\Gamma_{N_{1}}$ and $Y_{N}(x_{0})$. Furthermore, we have also shown an allowed parameter space for $\lambda_{L}-M_{DM}$ satisfying the relic abundance bound as well as the constraints from direct detection experiment of dark matter, XENON1T. 
As our model is compatible with baryogenesis studied in the IHDM desert, we are successfully able to show co-relation plot of dark matter mass ($M_{DM}$), RHN mass ($M_{2}$), lightest neutrino mass eigenvalue ($m_l$) and quartic coupling parameter ($\lambda_{5}$) with the latest observed value of BAU. We consider a particular range of quartic coupling, between $10^{-2}-5$, which is accountable for reproducing the observed baryon asymmetry of the Universe by the decay of $N_{2}$ with a mass in the range $ 10^{7}-5\times10^{8} $ GeV. We also calculate the light neutrino mass eigenvalues and check its consistency with the experimental bounds obtained from KamLAND-Zen by the neutrinoless double beta decay method. The correlation between the BAU result and $0\nu\beta\beta$ has a very constrained space in our work for both the mass ordering.   
From the synchronous study of $0\nu\beta\beta$ and baryogenesis, it is evident that both the observable are loosely co-related in our model. Moreover, the light neutrino mass eigenvalues obtained from this framework are more likely to satisfy the KamLAND-Zen limit for $m_{\beta\beta}$, and at the same time, they obey Planck limit for generating the observed BAU. From the co-relation plots between the various parameters and observed Planck limit of BAU, we can conclude that the NH is more preferable over the IH. \\
The significant conclusion we observe from our analysis is that the mass splitting, $\Delta M_{\eta^{\pm}} =\Delta M_{\eta^{0}_{I}} $ plays a vital role in the production of relic abundance via thermal production only. As, for thermal production of DM, we could generate relic for $\Delta M_{\eta^{\pm}} = \Delta M_{\eta^{0}_{I}} = 1 $ GeV but failed in the case of $\Delta M_{\eta^{\pm}}=\Delta M_{\eta^{0}_{I}} = 10 $ GeV for the same value of $\lambda_{L}=0.0001$, which therefore satisfies the LEP constraints \cite{Lundstrom:2008ai} as it rules out values of mass splitting greater than 8 GeV. This draws attention to how effective the mass splitting could be in the IHDM. It also motivates us to study the non-thermal production of dark matter. For non-thermal production of dark matter, we observe current relic abundance for the appropriate choice of decay width and coupling parameters with $\Delta M=1$ GeV and $\Delta M=10$ GeV. However, for $\Delta M_{\eta^{\pm}}=10$ GeV and $\Delta M_{\eta^{0}_{I}} = 0.01 $ GeV, we observe certain variations in the relic abundance curve. Thus, realising that the choice of mass splitting doesnot affect the relic abundance generated via non-thermal production unless they are equal. 
\section{Acknowledgement}
Authors would like to thank Debasish Borah of IIT Guwahati, India for his fruitful comments. The research work of PD and  MKD is supported by the Department of Science and Technology, Government of India, under the project grant EMR/2017/001436. LS would like to acknowledge Dibyendu Nanda of IITG, for his valuable help and discussions.
\bibliographystyle{utphys}
\bibliography{reff}
 \end{document}